# Physical Constraint Preserving Higher Order Finite Volume Schemes for Divergence-Free Astrophysical MHD and RMHD

By

Dinshaw S. Balsara[1,2], Deepak Bhoriya[1], Chetan Singh[3], Harish Kumar[3], Roger Käppeli[4] and Federico Gatti[4]

[1]Department of Physics and Astronomy, University of Notre Dame, IN, USA; dbalsara@nd.edu

[2]Applied and Computational Mathematics and Statistics, University of Notre Dame, IN, USA

[3]Department of Mathematics, Indian Institute of Technology, Delhi, India

[4]Seminar for Applied Mathematics (SAM), Department of Mathematics, ETH Zürich, CH-8092, Switzerland

**Abstract**

Higher order finite volume schemes for magnetohydrodynamics (MHD) and relativistic magnetohydrodynamics (RMHD) are very valuable because they allow us to carry out astrophysical simulations with very high accuracy. However, astrophysical problems sometimes have unusually large Mach numbers, exceptionally high Lorentz factors and very strong magnetic fields. All these effects cause higher order codes to become brittle and prone to code crashes. In this paper we document physical constraint preserving (PCP) methods for treating numerical MHD and RMHD. While unnecessary for standard problems, for stringent astrophysical problems these methods show their value. We describe higher order methods that allow divergence-free evolution of the magnetic field. We present a novel two-dimensional Riemann solver. This two-dimensional Riemann solver plays a key role in the design of PCP schemes for MHD and RMHD. We present a very simple PCP formulation and show how it is amalgamated with the evolution of face-centered magnetic fields. The methods presented here are time-explicit and do not add much to the computational cost. We show that the methods meet their design accuracies and work well on problems that would otherwise be considered too extreme for typical higher order Godunov methods of the type used in computational astrophysics.

## 1) Introduction

It is possible to claim that astrophysicists played a major role in the invention of higher order Godunov schemes (van Leer 1979, Colella and Woodward 1984, Woodward and Colella 1984). This was done with the express intent of inventing robust and accurate schemes for astrophysical hydrodynamics. These second order accurate methods were popular because they were based on three simple algorithmic elements:- First, one had to reconstruct the primal variables of the flow in order to get second, or better, order of spatial accuracy. Second, one had to invoke a Riemann solver for which knowledge of the eigenstructure could be beneficial. Third, one used a predictor-corrector (van Leer 1979, Colella and Woodward 1984, Colella 1985) or Runge-Kutta (Shu and Osher 1988) timestepping strategy to obtain second or better order accuracy in time. This winning plan was easy to adopt, resulting in the popularity of the above-mentioned papers for astrophysical hydrodynamics.

Owing to the presence of magnetic fields in astrophysical flows, it was only natural that astrophysicists wanted to invent higher order Godunov methods for magnetohydrodynamic (MHD) flows. Initial progress on understanding the MHD eigenstructure (Jeffrey and Taniuti 1964, Brio and Wu 1988, Zachary *et al*. 1994, Ryu and Jones 1995, Roe and Balsara 1996) was soon followed by the development of Riemann solvers for MHD (Cargo and Gallice 1997, Balsara 1998a, 1998b, Gurski 2004, Miyoshi and Kusano 2005, Li 2005). Inspired by the divergence-free discretizations of Yee (1966), and later on Brecht *et al*. (1981), DeVore (1991) and Evans and Hawley (1989), higher order Godunov schemes were designed which mimetically respected the divergence-free property of the magnetic field (Dai and Woodward 1998, Ryu *et al*. 1998, Balsara and Spicer 1999a). In time, higher order methods were developed for reconstructing the magnetic field in divergence-free fashion and with higher order accuracy (Balsara 2001a, 2004, 2009, Balsara *et al*. 2009, 2013, 2018, Balsara, Samantaray and Subramanian 2023) which allowed higher order MHD formulations to be achieved. The task of obtaining truly multidimensionally upwinded electric fields at edge centers that are needed for updating face-centered magnetic fields was also accomplished with the invention of multidimensional Riemann solvers (Balsara 2010, 2012a, 2014, 2015, Balsara and Nkonga 2017). Relativistic magnetohydrodynamics (RMHD) also plays an important role in computational astrophysics, with the result that many of these advances in MHD were paralleled in a few years by analogous advances in RMHD (Komissarov 1999, Balsara 2001b, Gammie *et al*. 2003, DelZanna *et al*. 2007, Tchekhovskoy *et al*. 2007, Mignone *et* *al*. 2006, 2009, Giacomazzo and Rezzolla 2006, Anton *et al*. 2010, Kim and Balsara, 2014, Balsara and Kim 2016, Cai *et al*. 2025) and several others.

Advances were also made in high accuracy methods (Harten *et al*. 1986, Shu and Osher 1988, 1989, Jiang and Shu 1996, Balsara and Shu 2000, Balsara, Garain and Shu 2016, Balsara *et al*. 2023, 2024a, 2024b, 2025), especially focusing on weighted essentially non-oscillatory (WENO) schemes but also including other efforts (McCorquodale and Colella 2013, Buchmüller *et al*. 2014). Those efforts at achieving higher order hydrodynamics also inspired an attempt to achieve higher order MHD and RMHD (Balsara 2009, Balsara *et al*. 2009, 2013, Seo and Ryu 2023, Balsara *et al*. 2025). However, astrophysical flows tend to have unusually large Mach numbers, exceptionally high Lorentz factors and very strong magnetic fields. These two effects tend to drive astrophysical flows to the point where pressures can become catastrophically negative resulting in a code crash. This has been the *most pressing problem in computational astrophysics and space physics*. Simply increasing the order of accuracy of an astrophysical code does not yield a solution to this problem. There were several early attempts to address this issue (Balsara and Spicer 1999b, Hu, Adams and Shu 2012, Balsara 2012b). A very early paper (Einfeldt *et al*. 1991) had shown that with the use of certain Riemann solvers of HLL (Harten Lax van Leer) and LLF (Local Lax Friedrich) type, positivity could be achieved in hydrodynamical simulations with first order schemes, however those results were not picked up or generalized for quite a while. In recent years, there has been an effort to generalize those results within the context of physical constraint preserving (PCP) methods (Wu and Shu 2018, 2020, Bhoriya *et al*. 2025, Balsara *et al*. 2025). Since higher order finite volume schemes which preserve the divergence of the magnetic field exactly are gaining popularity in computational astrophysics and space physics, the *twin goals* of this paper are:- 1) to show how one can *formulate PCP schemes* for divergence-free MHD and RMHD and 2) to also to *show how this can be done for higher order finite volume schemes*. This also provides us with an opportunity to show the simplicity of multidimensional Riemann solvers for MHD and RMHD as they have been formulated in the recent literature on finite difference schemes (Balsara *et al*. 2025), and how these should be extended to finite volume schemes. For the sake of completeness, it is also worth pointing out that finite difference WENO-based PCP schemes for divergence-free MHD and RMHD that go up to ninth order of accuracy have already been documented in Balsara *et al*. (2025). This paper, therefore, shows how those advances are to be extended to finite volume WENO-based schemes for divergence-free MHD and RMHD. This proves to be an entirely non-trivial extension because there are many design choices that apply to finite difference schemes that do not apply to finite volume schemes and vice versa.

The plan of this paper is as follows. Section 2 describes the systems of interest and their higher order spatial and temporal update. In Section 3 we provide a very brief synopsis of the recent multidimensional Riemann solver from Balsara *et al*. (2025) which plays a key role in the design of PCP schemes. In Section 4 we describe a baseline higher order finite volume scheme. This is useful because it gives us the essential structure on which the PCP ideas can be engrafted. In Section 5 we describe a low order scheme that is PCP and we also describe how it can be non-linearly hybridized with the higher order scheme to yield a method that is always PCP. Section 6 provides a step-by-step description of the PCP method described here. Section 7 shows that our methods meet their design accuracies. Section 8 focuses on extreme test problems, i.e. the type of test problem that cannot be done by a normal higher order Godunov scheme except by use of the PCP methods described here. Section 9 offers some conclusions.

**2) The MHD and RMHD Systems in Computational Astrophysics**

In this Section we very quickly describe the MHD and RMHD systems that are of interest in computational astrophysics. While we show the equations in their 2D forms just to save space; the implementation is fully 3D.

The MHD system for an ideal fluid can be written in a conservation form as

$$\frac{\partial}{\partial t}\begin{pmatrix} \rho \\ \rho v_x \\ \rho v_y \\ \rho v_z \\ \varepsilon \\ B_x \\ B_y \\ B_z \end{pmatrix} + \frac{\partial}{\partial x}\begin{pmatrix} \rho v_x \\ \rho v_x^2 + p + \mathbf{B}^2/8\pi - B_x^2/4\pi \\ \rho v_x v_y - B_x B_y/4\pi \\ \rho v_x v_z - B_x B_z/4\pi \\ (\varepsilon + p + \mathbf{B}^2/8\pi)v_x - B_x(\mathbf{v}\cdot\mathbf{B})/4\pi \\ 0 \\ (v_x B_y - v_y B_x) \\ -(v_z B_x - v_x B_z) \end{pmatrix} + \frac{\partial}{\partial y}\begin{pmatrix} \rho v_y \\ \rho v_y v_x - B_y B_x/4\pi \\ \rho v_y^2 + p + \mathbf{B}^2/8\pi - B_y^2/4\pi \\ \rho v_y v_z - B_y B_z/4\pi \\ (\varepsilon + p + \mathbf{B}^2/8\pi)v_y - B_y(\mathbf{v}\cdot\mathbf{B})/4\pi \\ -(v_x B_y - v_y B_x) \\ 0 \\ (v_y B_z - v_z B_y) \end{pmatrix} = 0$$

(2.1)

where $\varepsilon = \rho \mathbf{v}^2 / 2 + p/(\Gamma - 1) + \mathbf{B}^2 / 8\pi$ is the total energy density. The density is denoted by $\rho$; the pressure is written as "$p$"; the velocity components are given by $v_x$, $v_y$, $v_z$; and the magnetic field components by $B_x$, $B_y$, $B_z$. "$\Gamma$" denotes the ratio of specific heats. The vector of primitive variables that one seeks to extract from the vector of conserved variables is given by $(\rho, v_x, v_y, v_z, p, B_x, B_y, B_z)^T$. The usual challenge in MHD simulations has been that when the velocities or magnetic fields become too large, the pressure can become zero or negative. While co-evolving an entropy evolution equation can help, it also causes a loss of conservation when it is indeed invoked in a given zone.

The RMHD system is given by

$$\frac{\partial}{\partial t}\begin{pmatrix} \rho\gamma \\ m_x \\ m_y \\ m_z \\ \varepsilon \\ B_x \\ B_y \\ B_z \end{pmatrix} + \frac{\partial}{\partial x}\begin{pmatrix} \rho\gamma v_x \\ m_x v_x + p + \dfrac{\mathbf{b}^2}{2} - B_x b_x / \gamma \\ m_y v_x - B_x b_y / \gamma \\ m_z v_x - B_x b_z / \gamma \\ m_x \\ 0 \\ (v_x B_y - v_y B_x) \\ -(v_z B_x - v_x B_z) \end{pmatrix} + \frac{\partial}{\partial y}\begin{pmatrix} \rho\gamma v_y \\ m_x v_y - B_y b_x / \gamma \\ m_y v_y + p + \dfrac{\mathbf{b}^2}{2} - B_y b_y / \gamma \\ m_z v_y - B_y b_z / \gamma \\ m_y \\ -(v_x B_y - v_y B_x) \\ 0 \\ (v_y B_z - v_z B_y) \end{pmatrix} = 0 \qquad (2.2)$$

Here, $\rho$ is the density, $\gamma$ is the Lorentz factor, set $\{\varepsilon, m_x, m_y, m_z\}$ forms the four-momentum density, $\mathbf{v} = (v_x, v_y, v_z)^\top$ is the 3-velocity vector, $p$ is the pressure and $B_x, B_y, B_z$ are the components of the magnetic field. The speed of light is assumed to be unity; therefore, the Lorentz factor is defined by $\gamma = 1/\sqrt{1 - \mathbf{v}^2}$. The conserved quantities $m_i$ and $\varepsilon$ are given by

$$m_i = (\rho h \gamma^2 + \mathbf{B}^2) v_i - (\mathbf{v} \cdot \mathbf{B}) B_i \quad ; \quad \varepsilon = \rho h \gamma^2 - p + \frac{\mathbf{B}^2}{2} + \frac{\mathbf{v}^2 \mathbf{B}^2 - (\mathbf{v} \cdot \mathbf{B})^2}{2} \qquad (2.3)$$

where $h = 1 + \dfrac{\Gamma}{\Gamma - 1}\dfrac{p}{\rho}$ is the specific enthalpy and $\Gamma$ denotes the polytropic index. In the above, Latin indices range from 1 to 3. The components and magnitude of the covariant magnetic field are defined as a four vector with 4-components denoted by $b_\mu$ which should indeed be distinguished from the 3-vector $B_i$. The 4-vector representation of the magnetic field, $b_\mu$, and the 3-vector representation of the magnetic field, $B_i$, are related by the following relations:-

$$b_\mu = \gamma\left(\mathbf{v}\cdot\mathbf{B}, \frac{B_i}{\gamma^2} + (\mathbf{v}\cdot\mathbf{B})v_i\right) \quad ; \quad \mathbf{b}^2 = \frac{\mathbf{B}^2}{\gamma^2} + (\mathbf{v}\cdot\mathbf{B})^2 \tag{2.4}$$

Here the Greek indices range from 0 to 3 and the Roman indices range from 1 to 3. Cai *et al.* (2025) have presented a provably convergent and robust Newton-Raphson method to extract the primitives from the conservative variables. We use the same strategy here. Despite the guarantee that a physical set of conserved variables will always yield a unique physical set of primitive variables, the RMHD system still poses other problems. As with the MHD system, when the velocities or magnetic fields become too large, the system can produce unphysical pressures. For the RMHD system, this pathology can indeed prevent us from extracting the primitive variables from the system of conserved variables. But there is an even deeper question in RMHD:- What is the optimal choice of primitive variables? It turns out that routine interpolation of certain choices of variables can, for entirely numerical reasons, produce velocities that are superluminal. For that reason, Balsara and Kim (2016) suggested that the optimal primitive variables should be $\left(\rho, \gamma v_x, \gamma v_y, \gamma v_z, p, B_x, B_y, B_z\right)^T$. They showed that using this set of variables for the reconstruction always guarantees that a subluminal velocity can be extracted from the reconstructed polynomial. We make the same choice here. Eqn. (2.4) from Balsara and Kim (2016) shows that given $\gamma v_x, \gamma v_y, \gamma v_z$ it is very easy to extract the corresponding subliminal $v_x, v_y, v_z$.

It is easy to see that both eqn. (2.1) and eqn. (2.2) are in conservation form. As a result, the update of the conserved variables, i.e. the first five components in those equations, should be in flux-conservative fashion. The magnetic fields in both systems obey Faraday's law

$$\frac{\partial \mathbf{B}}{\partial t} + \nabla \times \mathbf{E} = 0. \tag{2.5}$$

Here the electric field vector is given constitutively by $\mathbf{E} = -\mathbf{v}\times\mathbf{B}$. Eqn. (2.5) satisfies an involution constraint which says that the magnetic field remains divergence-free for all time. It is now traditional in computational astrophysics and space physics to use mimetic discretizations,

Yee (1966), to ensure that the involution is preserved. Fig. 1 shows such a discretization where the facially-averaged magnetic field components are collocated at the face-centers and the edge-averaged electric fields that are used for their update are collocated at the edge-centers. The overbars on the magnetic field components and electric field components in Fig. 1 are intended to emphasize their facially-averaged and edge-averaged interpretations respectively.

It should be emphasized that high order of accuracy is achieved for the conserved variables by using high order reconstruction/interpolation. In recent years, weighted essentially non-oscillatory WENO techniques have become so well developed for reconstructing or interpolating zone-centered variables that they can be used in an almost "off-the-shelf" fashion; see Jiang and Shu (1996), Balsara and Shu (2000), Balsara, Garain and Shu (2016) and Balsara Samantaray and Subramanian (2023). The fluxes should also be integrated over the faces using high order quadrature; and multidimensional Gaussian quadrature formulae are easy to find on the internet. However, stability of the numerical fluxes is crucially important. It is achieved for the conserved variables only if the numerical flux that is used has a well-designed centered part as well as a part that provides numerical dissipation. Dumbser and Balsara (2016) have provided one such general purpose Riemann solver that does just that. It has been shown to work well for MHD and RMHD. Many more one-dimensional Riemann solvers for MHD and RMHD have been cited in the Introduction. The one-dimensional Riemann solver in Dumbser and Balsara (2016) does have one unique advantage which is that it can work with HLLE formulations which were originally shown by Einfeldt *et al*. (1991) to be conducive to preserving positivity of pressure.

Higher order WENO-based methods for the reconstruction of magnetic fields are also well-known; see Balsara (2009), Balsara *et al*. (2018), Balsara, Samantaray and Subramanian (2023). In Balsara, Samantaray and Subramanian (2023) a WENO-based method for the reconstruction of facial vector fields is presented that is referred to as almost divergence preserving (ADP). While it is not exactly divergence-free, it is so up to discretization error and it has the benefit of being very low cost. We use it in this work. Just as the numerical flux needs to have a dissipative component, the electric field that is used for the update of eqn. (2.5) also needs to have a suitable numerical diffusion. Such multidimensional Riemann solvers have been constructed in Balsara (2010, 2012a, 2014) and Balsara and Nkonga (2017). In Balsara *et al*. (2025), a very simple analysis of eqn. (2.5) has been done which yields a multidimensional Riemann solver. Since it will prove to be very useful in obtaining the PCP property, we will briefly describe it in the next Section.

**3) Recap of the 2D Riemann Solver**

Fig. 1 shows us that when obtaining the edge-centered electric fields we need a two-dimensional Riemann solver. Focusing on the z-component of the electric field in Fig. 1, we see that it would need to be upwinded in the x- and y-directions. Here we present such a 2D Riemann solver that has multidimensional upwinding. The detailed description is available in Balsara *et al.* (2025) but here we provide the reader with just enough detail to understand it at an intuitive level and enough knowledge to use it in a code. We focus only on the sub-system of the MHD/RMHD equations that governs the evolution of the magnetic fields. Eqn. (2.5) can be written in flux form as:-

$$\frac{\partial}{\partial t}\begin{pmatrix} B_x \\ B_y \\ B_z \end{pmatrix} + \frac{\partial}{\partial x}\begin{pmatrix} 0 \\ -E_z \\ E_y \end{pmatrix} + \frac{\partial}{\partial y}\begin{pmatrix} E_z \\ 0 \\ -E_x \end{pmatrix} + \frac{\partial}{\partial z}\begin{pmatrix} -E_y \\ E_x \\ 0 \end{pmatrix} = 0 \qquad (3.1)$$

In general, eqn. (3.1) will be a sub-portion of a larger PDE system. We just assume that the electric field is constitutively evaluated using other variables in the larger PDE system. We also assume that by using the information in 1D Riemann solvers at the faces that surround an edge, we can get extremal signal speeds at each edge of the mesh. Both these assumptions hold for MHD and RMHD. Fig. 2 shows four zones in the xy-plane that come together at the z-edge of a three-dimensional mesh. Since the mesh is viewed from the top in plan view, the z-edge is shown by the black dot and the four abutting zones are shown as four squares. The four incoming states have subscripts given by "RU" for right-upper; "LU" for left-upper; "LD" for left-down and "RD" for right-down. At each zone-center the z-component of the electric field can be evaluated, with the result that $E_{zRU}$, $E_{zLU}$, $E_{zLD}$ and $E_{zRD}$ form four of the inputs to the 2D Riemann problem shown in Fig. 2. The normal component of the magnetic field is continuous at the faces of the mesh, with the result that $B_{xD}$, $B_{xU}$, $B_{yR}$ and $B_{yL}$ form the other four inputs to the 2D Riemann problem shown in Fig. 2. Using the 1D speeds from the 1D Riemann solvers in the surrounding faces in Fig. 2, we can deduce that the 2D Riemann problem will have an extremal right-going speed given by $S_R$; an extremal left-going speed given by $S_L$, an extremal upward-going speed given by $S_U$ and an extremal down-going speed given by $S_D$. We will also have need for the maximal speed given by $S = \max(|S_R|, |S_L|, |S_U|, |S_D|)$ when discussing the LLF variant of the 2D Riemann solver that is presented here. These extremal speeds are shown in Fig. 3, and form a 2D wave model for the 2D Riemann solver.

Fig. 3 shows the same situation as Fig. 2. However, it shows the situation after the four incoming states start interacting with each other. Four one-dimensional Riemann problems, shown by dashed lines, develop between the four pairs of incoming states. The resolved states from the one-dimensional Riemann problems are shown by a superscript with a single star. The shaded region depicts the strongly interacting state that arises when the four one-dimensional Riemann problems interact with one another. The strongly interacting state is shown by a superscript with a double star. We want to find the z-component of the electric field in the strongly interacting state. This gives us the z-component of the electric field that overlies the z-edge, which is shown by the dot in this two-dimensional projection.

In Fig. 3 we see two x-directional Riemann problems which produce the states $B^*_{xU}, B^*_{yU}, E^*_{zU}$ and $B^*_{xD}, B^*_{yD}, E^*_{zD}$. Application of the HLL Riemann solver in the x-direction gives us

$$B^*_{xU} = B_{xU} \quad ; \quad B^*_{yU} = (S_R B_{yR} - S_L B_{yL})/(S_R - S_L) + (E_{zRU} - E_{zLU})/(S_R - S_L) \quad ;$$
$$E^*_{zU} = (S_R E_{zLU} - S_L E_{zRU})/(S_R - S_L) - S_R S_L (B_{yR} - B_{yL})/(S_R - S_L) \quad . \tag{3.2}$$

If we also want the LLF variant of the above equation, we get

$$B^*_{xU} = B_{xU} \quad ; \quad B^*_{yU} = (B_{yR} + B_{yL})/2 + (E_{zRU} - E_{zLU})/(2S) \quad ;$$
$$E^*_{zU} = (E_{zLU} + E_{zRU})/2 + S(B_{yR} - B_{yL})/2 \quad . \tag{3.3}$$

To obtain $B^*_{xD}, B^*_{yD}, E^*_{zD}$, we just have to set $U \to D$ in the above equations. In Fig. 3 we also see two y-directional Riemann problems which produce the states $B^*_{xR}, B^*_{yR}, E^*_{zR}$ and $B^*_{xL}, B^*_{yL}, E^*_{zL}$. Application of the HLL Riemann solver in the y-direction gives us

$$B_{xR}^* = (S_U B_{xU} - S_D B_{xD})/(S_U - S_D) - (E_{zRU} - E_{zRD})/(S_U - S_D) \quad ; \quad B_{yR}^* = B_{yR} \quad ;$$
$$E_{zR}^* = (S_U E_{zRD} - S_D E_{zRU})/(S_U - S_D) + S_U S_D (B_{xU} - B_{xD})/(S_U - S_D) \quad .$$
(3.4)

If we also want the LLF variant of the above equation, we get

$$B_{xR}^* = (B_{xU} + B_{xD})/2 - (E_{zRU} - E_{zRD})/(2S) \quad ; \quad B_{yR}^* = B_{yR} \quad ;$$
$$E_{zR}^* = (E_{zRD} + E_{zRU})/2 - S(B_{xU} - B_{xD})/2 \quad .$$
(3.5)

To obtain $B_{xL}^*, B_{yL}^*, E_{zL}^*$, we have just to set $R \to L$ in the above two equations. This paragraph has fully described all the states that surround the shaded strongly interacting state in Fig. 3.

Indeed, the shaded strongly interacting state in Fig. 3 is the thing that we seek. This state is shown with a double starred superscript in Fig. 3. The theory for 2D and 3D Riemann solvers was invented primarily to give us that state. Using the theory for 2D Riemann solvers that was developed in Balsara (2010, 2012a, 2014, 2015), we get two options for the resolved strongly interacting state $B_x^{**}, B_y^{**}, E_z^{**}$ in Fig. 3. The resolved magnetic fields, i.e. $B_x^{**}$ and $B_y^{**}$, are uniquely given by

$$B_x^{**} = (S_U B_{xU} - S_D B_{xD})/(S_U - S_D) + (E_{zLD} - E_{zLU} + E_{zRD} - E_{zRU})/(2(S_U - S_D)) \quad ;$$
$$B_y^{**} = (S_R B_{yR} - S_L B_{yL})/(S_R - S_L) + (-E_{zLD} - E_{zLU} + E_{zRD} + E_{zRU})/(2(S_R - S_L))$$
(3.6)

Depending on the choice of x-flux or y-flux we get

$$E_{z;1}^{**} = -(S_R + S_L) B_y^{**}/2 + (S_U (E_{zLD} + E_{zRD}) - S_D (E_{zLU} + E_{zRU}))/(2(S_U - S_D))$$
$$- S_U S_D (B_{xD} - B_{xU})/(S_U - S_D) + (S_R B_{yR} + S_L B_{yL})/2$$
(3.7)

or

$$E_{z;2}^{**} = (S_U + S_D) B_x^{**}/2 + (S_R (E_{zLD} + E_{zLU}) - S_L (E_{zRD} + E_{zRU}))/(2(S_R - S_L))$$
$$- (S_U B_{xU} + S_D B_{xD})/2 - S_R S_L (B_{yR} - B_{yL})/(S_R - S_L)$$
(3.8)

Since both choices are equally viable and entirely consistent, we choose $E_z^{**} = (E_{z;1}^{**} + E_{z;2}^{**})/2$. If we want the LLF variant of the above three equations, we get

$$B_x^{**} = (B_{xU} + B_{xD})/2 + (E_{zLD} - E_{zLU} + E_{zRD} - E_{zRU})/(4S) \quad ;$$
$$B_y^{**} = (B_{yR} + B_{yL})/2 + (E_{zRD} - E_{zLD} + E_{zRU} - E_{zLU})/(4S) \quad ; \qquad (3.9)$$
$$E_z^{**} = (E_{zRU} + E_{zLU} + E_{zLD} + E_{zRD})/4 + S[B_{xD} - B_{xU} + B_{yR} - B_{yL}]/2 \quad .$$

The last equation in 3.9 gives us a very useful insight. It tells us that just like a 1D Riemann solver, the electric field $E_z^{**}$ from the two dimensional Riemann solver that is suited for numerical work is also made up of two parts. The first part, $(E_{zRU} + E_{zLU} + E_{zLD} + E_{zRD})/4$, which we call the centered part, gives us consistency. In other words, it gives us the correct averaged electric field at the z-edge. The second part, $S[B_{xD} - B_{xU} + B_{yR} - B_{yL}]/2$, bears the dissipative contribution. This is the amount of dissipation that the 2D Riemann solver decrees should be present in the numerical electric field $E_z^{**}$ if the numerical magnetic field is to evolve stably on the mesh. Just like the 1D Riemann solver, the dissipation is proportional to the jumps in the solution. This completes our description of the states and the electric fields in Fig. 3 which depicts the 2D Riemann solver.

The last bit of insight that one needs in order to use the 2D Riemann solver consists of realizing that it is not just capable of giving us the subsonic state shown in Fig. 3, but it can also give us the other eight supersonic states that are shown in Fig. 4. The state that we will always be interested in is indeed the state that overlies the z-axis which is shown by the thick black dot in Figs. 3 and 4. Since the focus of this paper is utilitarian, we just document the result from Balsara *et al.* (2025) as:-

$$
\begin{aligned}
&if\ (S_L \geq 0\ and\ S_D \geq 0)\ then \\
&\quad E_z^{num} = E_{zLD}\ ;\ return\ ; \\
&elseif\ (S_R \leq 0\ and\ S_D \geq 0)\ then \\
&\quad E_z^{num} = E_{zRD}\ ;\ return\ ; \\
&elseif\ (S_R \leq 0\ and\ S_U \leq 0)\ then \\
&\quad E_z^{num} = E_{zRU}\ ;\ return\ ; \\
&elseif\ (S_L \geq 0\ and\ S_U \leq 0)\ then \\
&\quad E_z^{num} = E_{zLU}\ ;\ return\ ; \\
&elseif\ (S_L \geq 0)\ then \\
&\quad E_z^{num} = E_{zL}^*\ ;\ return\ ; \\
&elseif\ (S_R \leq 0)\ then \\
&\quad E_z^{num} = E_{zR}^*\ ;\ return\ ; \\
&elseif\ (S_D \geq 0)\ then \\
&\quad E_z^{num} = E_{zD}^*\ ;\ return\ ; \\
&elseif\ (S_U \leq 0)\ then \\
&\quad E_z^{num} = E_{zU}^*\ ;\ return\ ; \\
&else \\
&\quad E_z^{num} = \left(E_{z;1}^{**} + E_{z;2}^{**}\right)/2\ ;\ return\ ; \\
&endif
\end{aligned}
\qquad (3.10)
$$

For the LLF Riemann solver there are no supersonic states and the entire result is given by eqn. (3.9). This completes our description of the strategy for obtaining the numerical electric field for the induction equation. One can make cyclic variations of the formulae in this Section to obtain $E_x^{num}$ and $E_y^{num}$.

### 4) Higher Order Finite Volume Method for MHD and RMHD

All higher order finite volume schemes follow a similar plan. First, they obtain a higher order reconstruction of the solution that is valid at all points in the zone of interest. This is shown in Sub-section 4.1. Second, 1D Riemann solvers are invoked at zone faces to get the numerical fluxes at the faces; and 2D Riemann solvers are invoked at zone edges to get the numerical electric field components at the edges. This is discussed further in Sub-section 4.2. Third, an SSP-RK (Strong Stability Preserving-Runge Kutta) timestepping is invoked to get higher order temporal accuracy; as done in Sub-section 4.3 Of course, one can also build a higher order method that is based on a higher order predictor-corrector formulation, but it is easier to introduce PCP methods within the context of SSP-RK timestepping, which we do here. The rest of this Section describes each of the three parts.

**4.1) Higher Order Reconstruction of Facial and Zone-Centered Variables**

For the facial magnetic fields, the almost divergence preserving (ADP) reconstruction from Balsara, Samantaray and Subramanian (2023) is used. This step is done first. (The ADP reconstruction entails first reconstructing the magnetic field components two-dimensionally within the faces. Then this reconstruction is extended into the volume of the mesh. This ADP reconstruction yields a reconstruction of the vector field that is zero up to discretization error.) For the zone-centered variables, we start with the volume averaged vector of variables, which are denoted by $\bar{\mathbf{U}}_{i,j,k}$ in zone $(i, j, k)$; the overbar denotes volume averaging. We assume a uniform mesh with zones of size $\Delta x$, $\Delta y$ and $\Delta z$ in each of the three directions. We wish to take a timestep of size $\Delta t$ that is within the CFL limit. Because the target scheme is PCP, it is very advantageous to reconstruct the primitive variables with high order accuracy. However, in order to do that, we have to carry out the following three steps:- 1) Obtain zone-centered point values for the conserved variables. 2) Obtain therefrom the corresponding zone-centered point values for the primitive variables. 3) Then obtain zone-averaged primitive variables. These can then be reconstructed with high order of accuracy. At second order, this is easily done. However, at third and higher orders, the process becomes more intricate. The next paragraph describes how this is done at third and fourth orders; the paragraph after that describes how this is done at fifth and sixth orders.

At fourth order, McCorquodale and Colella (2011) showed how this three-step procedure is done, and we draw inspiration from eqns. (12) and (16) of that paper. Starting from a mesh function that is made of zone-averaged conserved variables $\bar{\mathbf{U}}_{i,j,k}$, we can obtain the corresponding point value $\mathbf{U}_{i,j,k}$ at the center of zone $(i, j, k)$. The problem is that the process entails subtracting off higher order derivatives from $\bar{\mathbf{U}}_{i,j,k}$, and if those higher order derivatives are obtained from a discontinuous solution, it can seriously damage the extraction of point values. Therefore, it is possible to define a flattener function, $\eta_{i,j,k}$, within each zone such that $\eta_{i,j,k} = 0$ for smooth flow and $\eta_{i,j,k} \to 1$ for non-smooth flow. Let us also define the function $\phi_{i,j,k} \equiv 1 - \eta_{i,j,k}$ within each zone such that $\phi_{i,j,k} = 1$ for smooth flow and $\phi_{i,j,k} \to 0$ when the flow becomes increasingly non-smooth. The description of such flattener functions for Euler flow is given in Colella and Woodward (1994), for MHD flows the flattener function is described in Balsara (2012b) and for RMHD flows it is presented in Balsara and Kim (2016). (The flattener function, which yields the coefficients $\eta_{i,j,k}$ and $\phi_{i,j,k}$ for RMHD, is defined by eqns. (2.8) and (2.9) in Balsara and Kim (2016).) If the flow is smooth, i.e. if $\phi_{i,j,k} = 1$, third or fourth order accuracy (depending on what is desired) can be ensured. In the fourth order limit we can assert

$$\mathbf{U}_{i,j,k} = \overline{\mathbf{U}}_{i,j,k} - \frac{1}{12}\phi_{i,j,k}\left[u_{xx} + u_{yy} + u_{zz}\right] \quad \text{with} \quad u_{xx} = \left(\overline{\mathbf{U}}_{i+1,j,k} - 2\overline{\mathbf{U}}_{i,j,k} + \overline{\mathbf{U}}_{i-1,j,k}\right)/2 \;;$$
$$u_{yy} = \left(\overline{\mathbf{U}}_{i,j+1,k} - 2\overline{\mathbf{U}}_{i,j,k} + \overline{\mathbf{U}}_{i,j-1,k}\right)/2 \quad;\quad u_{zz} = \left(\overline{\mathbf{U}}_{i,j,k+1} - 2\overline{\mathbf{U}}_{i,j,k} + \overline{\mathbf{U}}_{i,j,k-1}\right)/2$$
(4.1)

In other words, when the flow is sufficiently smooth, eqn. (4.1) will start with the volume averaged quantities (i.e. the ones with overbars) and give us a fourth order accurate value for $\mathbf{U}_{i,j,k}$, the vector of conserved variables which is defined pointwise at the zone center. Using standard root solver techniques that are available in the literature (for RMHD we recommend Cai *et al.* 2025), we obtain $\mathbf{V}_{i,j,k}$, the vector of primitive variables defined pointwise at the zone centers of all the zones of the mesh. From these zone-centered point values for the primitive variables, we can obtain zone averaged values of the primitive variables, which we denote with an overbar. Again, at third and fourth order accuracy we obtain

$$\overline{\mathbf{V}}_{i,j,k} = \mathbf{V}_{i,j,k} + \frac{1}{12}\phi_{i,j,k}\left[v_{xx} + v_{yy} + v_{zz}\right] \quad \text{with} \quad v_{xx} = \left(\mathbf{V}_{i+1,j,k} - 2\mathbf{V}_{i,j,k} + \mathbf{V}_{i-1,j,k}\right)/2 \;;$$
$$v_{yy} = \left(\mathbf{V}_{i,j+1,k} - 2\mathbf{V}_{i,j,k} + \mathbf{V}_{i,j-1,k}\right)/2 \quad;\quad v_{zz} = \left(\mathbf{V}_{i,j,k+1} - 2\mathbf{V}_{i,j,k} + \mathbf{V}_{i,j,k-1}\right)/2$$
(4.2)

Since the goal is to arrive at a PCP method, we suggest that it is better to use reconstruction formulae on the zone averaged primitive variables in eqn. (4.2) than to use reconstruction on the conserved variables in eqn. (4.1). For that reason, the zone averaged primitive variables $\overline{\mathbf{V}}_{i,j,k}$ can be reconstructed with fourth order of accuracy using a multidimensional WENO reconstruction scheme like the one described in the supplement of Balsara, Samantaray and Subramanian (2023). (We have suitably documented this reconstruction strategy to 3D and up to fifth and sixth order in the above reference, and it has also been extended to even higher orders by us.) As a result, within each zone we have a third or fourth order accurate reconstructed function for the vector of primitive variables given by $\hat{\mathbf{V}}_{i,j,k}(x,y,z)$. The advantage of reconstructing the primitive variables is that one can immediately test whether the reconstructed variable is within the PCP domain. If it is not, then the higher order modes of the reconstructed variable can be reduced to ensure that the interpolating function remains within the PCP domain.

While McCorquodale and Colella (2011) only went up to fourth order, the same philosophy can be used in conjunction with WENO reconstruction (see the supplement of Balsara, Samantaray and Subramanian 2023) to go to even higher orders. Therefore, for a fifth or sixth order scheme, the equation that is analogous to eqn. (4.1) is given by

$$\mathbf{U}_{i,j,k} = \overline{\mathbf{U}}_{i,j,k} - \frac{1}{12}\phi_{i,j,k}\left[u_{xx} + u_{yy} + u_{zz}\right] + \frac{3}{560}\phi_{i,j,k}\left[u_{xxxx} + u_{yyyy} + u_{zzzz}\right] + \frac{1}{144}\phi_{i,j,k}\left[u_{xxyy} + u_{yyzz} + u_{xxzz}\right]$$

with

$$u_{xx} = (-74\overline{\mathbf{U}}_{i,j,k} + 40\overline{\mathbf{U}}_{i-1,j,k} - 3\overline{\mathbf{U}}_{i-2,j,k} + 40\overline{\mathbf{U}}_{i+1,j,k} - 3\overline{\mathbf{U}}_{i+2,j,k})/56;$$
$$u_{yy} = (-74\overline{\mathbf{U}}_{i,j,k} + 40\overline{\mathbf{U}}_{i,j-1,k} - 3\overline{\mathbf{U}}_{i,j-2,k} + 40\overline{\mathbf{U}}_{i,j+1,k} - 3\overline{\mathbf{U}}_{i,j+2,k})/56;$$
$$u_{zz} = (-74\overline{\mathbf{U}}_{i,j,k} + 40\overline{\mathbf{U}}_{i,j,k-1} - 3\overline{\mathbf{U}}_{i,j,k-2} + 40\overline{\mathbf{U}}_{i,j,k+1} - 3\overline{\mathbf{U}}_{i,j,k+2})/56;$$
$$u_{xxxx} = (6\overline{\mathbf{U}}_{i,j,k} - 4\overline{\mathbf{U}}_{i-1,j,k} + \overline{\mathbf{U}}_{i-2,j,k} - 4\overline{\mathbf{U}}_{i+1,j,k} + \overline{\mathbf{U}}_{i+2,j,k})/24;$$
$$u_{yyyy} = (6\overline{\mathbf{U}}_{i,j,k} - 4\overline{\mathbf{U}}_{i,j-1,k} + \overline{\mathbf{U}}_{i,j-2,k} - 4\overline{\mathbf{U}}_{i,j+1,k} + \overline{\mathbf{U}}_{i,j+2,k})/24;$$
$$u_{zzzz} = (6\overline{\mathbf{U}}_{i,j,k} - 4\overline{\mathbf{U}}_{i,j,k-1} + \overline{\mathbf{U}}_{i,j,k-2} - 4\overline{\mathbf{U}}_{i,j,k+1} + \overline{\mathbf{U}}_{i,j,k+2})/24;$$
$$u_{xxyy} = (4\overline{\mathbf{U}}_{i,j,k} - 2\overline{\mathbf{U}}_{i,j-1,k} - 2\overline{\mathbf{U}}_{i,j+1,k} - 2\overline{\mathbf{U}}_{i-1,j,k} + \overline{\mathbf{U}}_{i-1,j-1,k} + \overline{\mathbf{U}}_{i-1,j+1,k} - 2\overline{\mathbf{U}}_{i+1,j,k} + \overline{\mathbf{U}}_{i+1,j-1,k} + \overline{\mathbf{U}}_{i+1,j+1,k})/4;$$
$$u_{yyzz} = (4\overline{\mathbf{U}}_{i,j,k} - 2\overline{\mathbf{U}}_{i,j,k-1} - 2\overline{\mathbf{U}}_{i,j,k+1} - 2\overline{\mathbf{U}}_{i,j-1,k} + \overline{\mathbf{U}}_{i,j-1,k-1} + \overline{\mathbf{U}}_{i,j-1,k+1} - 2\overline{\mathbf{U}}_{i,j+1,k} + \overline{\mathbf{U}}_{i,j+1,k-1} + \overline{\mathbf{U}}_{i,j+1,k+1})/4;$$
$$u_{xxzz} = (4\overline{\mathbf{U}}_{i,j,k} - 2\overline{\mathbf{U}}_{i,j,k-1} - 2\overline{\mathbf{U}}_{i,j,k+1} - 2\overline{\mathbf{U}}_{i-1,j,k} + \overline{\mathbf{U}}_{i-1,j,k-1} + \overline{\mathbf{U}}_{i-1,j,k+1} - 2\overline{\mathbf{U}}_{i+1,j,k} + \overline{\mathbf{U}}_{i+1,j,k-1} + \overline{\mathbf{U}}_{i+1,j,k+1})/4;$$

(4.3)

Eqn. (4.3) at fifth and sixth orders is analogous to eqn. (4.1) at fourth order. It is worthwhile noting that the moments $u_{xx}$, $u_{yy}$, $u_{zz}$, $u_{xxxx}$, $u_{yyyy}$, $u_{zzzz}$, $u_{xxyy}$, $u_{yyzz}$ and $u_{xxzz}$ in the above formula are obtained from a high order reconstruction. As before, within each zone we can use the zone-centered conserved variable $\mathbf{U}_{i,j,k}$ to build the zone-centered point values for the vector of primitives $\mathbf{V}_{i,j,k}$. At fifth and sixth orders, the transcription that is analogous to eqn. (4.2) is given via a high order interpolation and is given by

$$\bar{\mathbf{V}}_{i,j,k} = \mathbf{V}_{i,j,k} + \frac{1}{12}\phi_{i,j,k}\left[v_{xx} + v_{yy} + v_{zz}\right] - \frac{3}{560}\phi_{i,j,k}\left[v_{xxxx} + v_{yyyy} + v_{zzzz}\right] - \frac{1}{144}\phi_{i,j,k}\left[v_{xxyy} + v_{yyzz} + v_{xxzz}\right]$$

with

$$v_{xx} = (-346\mathbf{V}_{i,j,k} - 14\mathbf{V}_{i,j,k-1} - 14\mathbf{V}_{i,j,k+1} - 14\mathbf{V}_{i,j-1,k} - 14\mathbf{V}_{i,j+1,k} + 184\mathbf{V}_{i-1,j,k} + 7\mathbf{V}_{i-1,j,k-1} + 7\mathbf{V}_{i-1,j,k+1}$$
$$+ 7\mathbf{V}_{i-1,j-1,k} + 7\mathbf{V}_{i-1,j+1,k} - 11\mathbf{V}_{i-2,j,k} + 184\mathbf{V}_{i+1,j,k} + 7\mathbf{V}_{i+1,j,k-1} + 7\mathbf{V}_{i+1,j,k+1} + 7\mathbf{V}_{i+1,j-1,k} + 7\mathbf{V}_{i+1,j+1,k}$$
$$- 11\mathbf{V}_{i+2,j,k})/336$$

$$v_{yy} = (-346\mathbf{V}_{i,j,k} - 14\mathbf{V}_{i,j,k-1} - 14\mathbf{V}_{i,j,k+1} + 184\mathbf{V}_{i,j-1,k} + 7\mathbf{V}_{i,j-1,k-1} + 7\mathbf{V}_{i,j-1,k+1} - 11\mathbf{V}_{i,j-2,k} + 184\mathbf{V}_{i,j+1,k}$$
$$+ 7\mathbf{V}_{i,j+1,k-1} + 7\mathbf{V}_{i,j+1,k+1} - 11\mathbf{V}_{i,j+2,k} - 14\mathbf{V}_{i-1,j,k} + 7\mathbf{V}_{i-1,j-1,k} + 7\mathbf{V}_{i-1,j+1,k} - 14\mathbf{V}_{i+1,j,k} + 7\mathbf{V}_{i+1,j-1,k}$$
$$+ 7\mathbf{V}_{i+1,j+1,k})/336$$

$$v_{zz} = (-346\mathbf{V}_{i,j,k} + 184\mathbf{V}_{i,j,k-1} - 11\mathbf{V}_{i,j,k-2} + 184\mathbf{V}_{i,j,k+1} - 11\mathbf{V}_{i,j,k+2} - 14\mathbf{V}_{i,j-1,k} + 7\mathbf{V}_{i,j-1,k-1} + 7\mathbf{V}_{i,j-1,k+1}$$
$$- 14\mathbf{V}_{i,j+1,k} + 7\mathbf{V}_{i,j+1,k-1} + 7\mathbf{V}_{i,j+1,k+1} - 14\mathbf{V}_{i-1,j,k} + 7\mathbf{V}_{i-1,j,k-1} + 7\mathbf{V}_{i-1,j,k+1} - 14\mathbf{V}_{i+1,j,k} + 7\mathbf{V}_{i+1,j,k-1}$$
$$+ 7\mathbf{V}_{i+1,j,k+1})/336$$

$$v_{xxxx} = (6\mathbf{V}_{i,j,k} - 4\mathbf{V}_{i-1,j,k} + \mathbf{V}_{i-2,j,k} - 4\mathbf{V}_{i+1,j,k} + \mathbf{V}_{i+2,j,k})/24$$
$$v_{yyyy} = (6\mathbf{V}_{i,j,k} - 4\mathbf{V}_{i,j-1,k} + \mathbf{V}_{i,j-2,k} - 4\mathbf{V}_{i,j+1,k} + \mathbf{V}_{i,j+2,k})/24$$
$$v_{zzzz} = (6\mathbf{V}_{i,j,k} - 4\mathbf{V}_{i,j,k-1} + \mathbf{V}_{i,j,k-2} - 4\mathbf{V}_{i,j,k+1} + \mathbf{V}_{i,j,k+2})/24$$
$$v_{xxyy} = (4\mathbf{V}_{i,j,k} - 2\mathbf{V}_{i,j-1,k} - 2\mathbf{V}_{i,j+1,k} - 2\mathbf{V}_{i-1,j,k} + \mathbf{V}_{i-1,j-1,k} + \mathbf{V}_{i-1,j+1,k} - 2\mathbf{V}_{i+1,j,k} + \mathbf{V}_{i+1,j-1,k} + \mathbf{V}_{i+1,j+1,k})/4$$
$$v_{yyzz} = (4\mathbf{V}_{i,j,k} - 2\mathbf{V}_{i,j,k-1} - 2\mathbf{V}_{i,j,k+1} - 2\mathbf{V}_{i,j-1,k} + \mathbf{V}_{i,j-1,k-1} + \mathbf{V}_{i,j-1,k+1} - 2\mathbf{V}_{i,j+1,k} + \mathbf{V}_{i,j+1,k-1} + \mathbf{V}_{i,j+1,k+1})/4$$
$$v_{xxzz} = (4\mathbf{V}_{i,j,k} - 2\mathbf{V}_{i,j,k-1} - 2\mathbf{V}_{i,j,k+1} - 2\mathbf{V}_{i-1,j,k} + \mathbf{V}_{i-1,j,k-1} + \mathbf{V}_{i-1,j,k+1} - 2\mathbf{V}_{i+1,j,k} + \mathbf{V}_{i+1,j,k-1} + \mathbf{V}_{i+1,j,k+1})/4$$

(4.4)

These zone-averaged primitive variables can be reconstructed with fifth or sixth order of accuracy using a multidimensional WENO reconstruction scheme like the one described in the supplement of Balsara, Samantaray and Subramanian (2023). Therefore, within each zone we have a third, fourth, fifth or sixth order accurate reconstructed function for the vector of primitive variables given by $\hat{\mathbf{V}}_{i,j,k}(x,y,z)$.

In the most stringent of situations, these reconstructed functions $\hat{\mathbf{V}}_{i,j,k}(x,y,z)$ may not be PCP at all locations of interest in the $(i,j,k)$ zone. If we find ourselves in such a situation, we can always find a $\kappa_{i,j,k} \in [0,1]$ such that a reset

$$\hat{\mathbf{V}}_{i,j,k}(x,y,z) \rightarrow \kappa_{i,j,k}\hat{\mathbf{V}}_{i,j,k}(x,y,z) + (1-\kappa_{i,j,k})\bar{\mathbf{V}}_{i,j,k} \qquad (4.5)$$

will bring the reconstructed functions within the PCP domain. The same plan can be used for the facial magnetic fields so that if we consider the higher order reconstructed x-component of the facial magnetic field to be $\hat{B}_{x;i+1/2,j,k}(y,z)$ and the mean value of the magnetic field in the same x-face to be $\bar{B}_{x;i+1/2,j,k}$, then we can write an equation that is analogous to eqn. (4.5) as

$$\hat{B}_{x;i+1/2,j,k}(y,z) \rightarrow \min(\kappa_{i,j,k},\kappa_{i+1,j,k})\hat{B}_{x;i+1/2,j,k}(y,z) + (1-\min(\kappa_{i,j,k},\kappa_{i+1,j,k}))\bar{B}_{x;i+1/2,j,k} \qquad (4.6)$$

This Sub-section has shown us that if we start with $\bar{\mathbf{U}}_{i,j,k}$ which is PCP in the zone $(i,j,k)$ then we can always obtain a reconstructed function for the vector of primitive variables given by $\hat{\mathbf{V}}_{i,j,k}(x,y,z)$ which retains the PCP property. Since these primitive variables are handed to the Riemann solvers, we guarantee that the variables that are fed to all the Riemann solvers are physical.

**4.2) Invoking 1D and Multidimensional Riemann Solvers**

Fig. 5 shows the arrangement of the spatial nodes in the fourth order accurate RK-WENO algorithm for two space dimensions. The nodes within four abutting spatial zones are shown by the black dots. At fourth order, one-dimensional Gaussian quadrature requires the use of three quadrature points; we, therefore, have three nodes within each face of Fig. 5. When two nodes from opposing faces abut one another, we invoke a one-dimensional Riemann solver. The red, double-sided arrows indicate the application of 1D Riemann solvers at the nodal points in the x-direction. The blue, double-sided arrows indicate the application of 1D Riemann solvers at the nodal points in the y-direction. The two input states that go into each 1D Riemann solver must indeed be PCP. Therefore, eqn. (4.5) should certainly be used to make sure that whichever interpolant is used within each zone produces PCP values at all the nodal points within that zone.

At each zone center, we should also construct the pointwise electric field components. This is easily done because we have all the primitive variables at each zone center. These zone-centered electric fields are then interpolated in 3D to high order. At each edge, the electric fields from the four abutting zones are then interpolated to that edge, as shown in Fig. 2. This then forms the centered part of the multidimensional Riemann solver. The centered part gives us consistency in the Riemann solver. The facial magnetic field components have also been reconstructed within the faces of the mesh to high order in the ADP reconstruction step. As a result, the jumps in the facial magnetic fields form the dissipation terms in the multidimensional Riemann solver. The dissipation terms give us numerical stabilization. (The reader may want to circle back to the discussion around eqn. (3.9) in order to recapitulate these ideas about a centered part and a dissipation part in the multidimensional Riemann solver.) The green dashed square at the right-upper vertex of zone *(i,j)* in Fig. 5 indicates the application of a 2D Riemann solver at the vertices of the mesh. This gives us the z-component of the electric field. Fig. 5 depicts a 2D mesh just to make it easier to visualize the concepts; however, the entire discussion extends to full three dimensions.

In full 3D, the time rate of change for the conserved variables can be written up to higher order as

$$\partial_t \overline{\mathbf{U}}_{i,j,k} = -\frac{1}{\Delta x}\left(\overline{\mathbf{F}}^{HO}_{i+1/2,j,k} - \overline{\mathbf{F}}^{HO}_{i-1/2,j,k}\right) - \frac{1}{\Delta y}\left(\overline{\mathbf{G}}^{HO}_{i,j+1/2,k} - \overline{\mathbf{G}}^{HO}_{i,j-1/2,k}\right) - \frac{1}{\Delta z}\left(\overline{\mathbf{H}}^{HO}_{i,j,k+1/2} - \overline{\mathbf{H}}^{HO}_{i,j,k-1/2}\right) \quad (4.7)$$

The superscripts "HO" for the fluxes indicate that these are high order numerical quadratures of the fluxes evaluated at the faces of the mesh. The overbars for the fluxes indicate that these are facially averaged fluxes. Likewise, the time rate of change for the facially-averaged magnetic field variables can be written up to higher order as

$$\partial_t \overline{B}_{x;\,i+1/2,j,k} = -\frac{1}{\Delta y \Delta z}\left(\Delta z \overline{E}^{HO}_{z;\,i+1/2,j+1/2,k} - \Delta z \overline{E}^{HO}_{z;\,i+1/2,j-1/2,k} + \Delta y \overline{E}^{HO}_{y;\,i+1/2,j,k-1/2} - \Delta y \overline{E}^{HO}_{y;\,i+1/2,j,k+1/2}\right) \quad (4.8)$$

$$\partial_t \overline{B}_{y;\,i,j-1/2,k} = -\frac{1}{\Delta x \Delta z}\left(\Delta x \overline{E}^{HO}_{x;\,i,j-1/2,k+1/2} - \Delta x \overline{E}^{HO}_{x;\,i,j-1/2,k-1/2} + \Delta z \overline{E}^{HO}_{z;\,i-1/2,j-1/2,k} - \Delta z \overline{E}^{HO}_{z;\,i+1/2,j-1/2,k}\right) \quad (4.9)$$

$$\partial_t \overline{B}_{z;\,i,j,k+1/2} = -\frac{1}{\Delta x \Delta y}\left(\Delta x \overline{E}^{HO}_{x;\,i,j-1/2,k+1/2} - \Delta x \overline{E}^{HO}_{x;\,i,j+1/2,k+1/2} + \Delta y \overline{E}^{HO}_{y;\,i+1/2,j,k+1/2} - \Delta y \overline{E}^{HO}_{y;\,i-1/2,j,k+1/2}\right) \quad (4.10)$$

As before, the superscripts "HO" for the electric fields indicate that these are high order numerical quadratures of the electric fields evaluated along the edges of the mesh. The overbars for the electric fields indicate that these are edge-averaged electric fields. The above four equations tell us how the time rate of change of the primal variables, i.e. the zone-averaged fluid variables and the facially averaged magnetic field components, are updated with high order of accuracy.

**4.3) SSP-RK Timestepping, Written Differently**

Eqns. (4.7) to (4.10) can now be integrated into a higher order SSP-RK timestepping (Shu and Osher 1988, Spiteri and Ruuth 2002, 2003). However, for the sake of understanding the steps that are to come later, we write this update a little differently in the narrative that follows. For the second order SSP-RK update we write

$$\begin{aligned}
\overline{\mathbf{U}}_i^{(1)} &= \overline{\mathbf{U}}_i^{(n)} + \Delta t\, \partial_t \overline{\mathbf{U}}_i^{(n)} \\
\overline{\mathbf{U}}_i^{(n+1)} &= \frac{1}{2}\overline{\mathbf{U}}_i^{(n)} + \frac{1}{2}\left(\overline{\mathbf{U}}_i^{(1)} + \Delta t\, \partial_t \overline{\mathbf{U}}_i^{(1)}\right)
\end{aligned} \tag{4.11}$$

For the third order, the similar style of update can be demonstrated as

$$\begin{aligned}
\overline{\mathbf{U}}_i^{(1)} &= \overline{\mathbf{U}}_i^{(n)} + \Delta t\, \partial_t \overline{\mathbf{U}}_i^{(n)} \\
\overline{\mathbf{U}}_i^{(2)} &= \frac{3}{4}\overline{\mathbf{U}}_i^{(n)} + \frac{1}{4}\left(\overline{\mathbf{U}}_i^{(1)} + \Delta t\, \partial_t \overline{\mathbf{U}}_i^{(1)}\right) \\
\overline{\mathbf{U}}_i^{(n+1)} &= \frac{1}{3}\overline{\mathbf{U}}_i^{(n)} + \frac{2}{3}\left(\overline{\mathbf{U}}_i^{(2)} + \Delta t\, \partial_t \overline{\mathbf{U}}_i^{(2)}\right)
\end{aligned} \tag{4.12}$$

A similar structure can be shown for the five-stage fourth order SSP-RK update as follows

$$\begin{aligned}
\overline{\mathbf{U}}_i^{(1)} &= 0.608247773428110\,\overline{\mathbf{U}}_i^{(n)} + 0.391752226571890\left(\overline{\mathbf{U}}_i^{(n)} + \Delta t\, \partial_t \overline{\mathbf{U}}_i^{(n)}\right) \\
\overline{\mathbf{U}}_i^{(2)} &= 0.444370493651235\,\overline{\mathbf{U}}_i^{(n)} + 0.187218913298394\,\overline{\mathbf{U}}_i^{(1)} + 0.368410593050371\left(\overline{\mathbf{U}}_i^{(1)} + \Delta t\, \partial_t \overline{\mathbf{U}}_i^{(1)}\right) \\
\overline{\mathbf{U}}_i^{(3)} &= 0.620101851488403\,\overline{\mathbf{U}}_i^{(n)} + 0.128006374239903\,\overline{\mathbf{U}}_i^{(2)} + 0.251891774271694\left(\overline{\mathbf{U}}_i^{(2)} + \Delta t\, \partial_t \overline{\mathbf{U}}_i^{(2)}\right) \\
\overline{\mathbf{U}}_i^{(4)} &= 0.178079954393132\,\overline{\mathbf{U}}_i^{(n)} + 0.276945295378347\,\overline{\mathbf{U}}_i^{(3)} + 0.544974750228521\left(\overline{\mathbf{U}}_i^{(3)} + \Delta t\, \partial_t \overline{\mathbf{U}}_i^{(3)}\right) \\
\overline{\mathbf{U}}_i^{(n+1)} &= 0.517231671970585\,\overline{\mathbf{U}}_i^{(2)} + \left\{0.032367241859857\,\overline{\mathbf{U}}_i^{(3)} + 0.063692468666290\left(\overline{\mathbf{U}}_i^{(3)} + \Delta t\, \partial_t \overline{\mathbf{U}}_i^{(3)}\right)\right\} \\
&\quad + \left\{0.160701134266363\,\overline{\mathbf{U}}_i^{(4)} + 0.226007483236906\left(\overline{\mathbf{U}}_i^{(4)} + \Delta t\, \partial_t \overline{\mathbf{U}}_i^{(4)}\right)\right\}
\end{aligned} \tag{4.13}$$

The common feature in eqns. (4.11) to (4.13) is that each stage has been written in the form of a forward Euler update. In the next Section we will show that we can make a forward Euler update PCP by nonlinearly hybridizing any high order code with a first order code. In doing so, we will have shown that every stage in an SSP-RK update can be written in terms of a convex combination of states, each of which is PCP. As a result, the entire scheme will be made PCP.

It is also worth noting that several problems in computational astrophysics do not have unusually large Mach numbers, neither do they have exceptionally high Lorentz factors and nor do they have ultra-strong magnetic fields. For such problems, even the algorithm that is described up to this point would yield a high order algorithm that does not need PCP enforcement. Thus the PCP property is an add-on step to help out in very stringent astrophysical and space physics simulations.

**5) A First Order PCP Scheme that can be Hybridized with the Previous Section**

When equations (4.7) to (4.10) are integrated into one of the three SSP-RK time-stepping schemes shown in eqns. (4.11), (4.12) or (4.13), the result is a scheme with higher order accuracy in space and time. Because of eqns. (4.5) and (4.6) we can guarantee that the method will never provide inputs to a Riemann solver that are not PCP. This ensures that the Riemann solvers will always be able to return meaningful higher order fluxes and electric fields. The problem is that this is still not sufficient to ensure that the final update for each stage in the multi-stage Runge-Kutta timestepping remains PCP. Our first task, which we take up in Sub-Section 5.1, is to describe a first order scheme that unconditionally stays within the PCP region. Our second task, which we take up in Sub-Section 5.2, is to describe how such a scheme can be integrated into a modified version of eqns. (4.7) to (4.10) in such a way that for troubled zones, and only for troubled zones, we gradually resort to the lower order scheme.

**5.1) A First Order PCP Scheme for Divergence-Free MHD and RMHD**

Let us first start with an interesting observation. A first order, zone-centered scheme for MHD and RMHD that is updated with fluxes that are obtained with an HLL or LLF Riemann solver will indeed be PCP. See Gurski (2004) for a proof of this. No reconstruction of any sort is needed for such a scheme that is first order in space and time; as a result, the scheme is extremely inexpensive. This scheme only uses a single call to a one-dimensional HLL or LLF Riemann solver at each face, thus guaranteeing that it is very lightweight compared to any higher order scheme. In other words, we have carried out extensive numerical experimentation to show that for such a scheme, if all the variables in all the zones are PCP at the beginning of a forward Euler timestep, then they will remain PCP at the end of the timestep. This is a good observation because it gives us something to build on.

For any first order scheme we can always average facial magnetic fields to obtain zone-centered magnetic fields. Because the facial magnetic fields are our primal magnetic fields, those fields will have to be updated using first order accurate edge-centered electric fields, as described in eqns. (3.6) to (3.8), or more simply in eqn. (3.9). As in the previous paragraph, this only requires one, extremely light weight evaluation of the 2D Riemann solver at each edge and eqn. (3.9) shows that this evaluation is computationally very inexpensive. Even for the most stringent of problems, such a scheme will remain PCP in most of the zones. There will be a rare few zones where the averaging of the facial magnetic fields to the zone centers will indeed destroy the PCP property. To get a bullet-proof scheme, we have to find a way to cure this problem. (The subsequent non-linear hybridization of such a low order PCP scheme with a higher order scheme that is not PCP will be described in the next Sub-Section.)

Let us re-articulate the previous paragraph in mathematical terms to make it more comprehensible. We start with $\bar{\mathbf{U}}_{i,j,k}^{n;LO}$ where the sixth, seventh and eighth components of $\bar{\mathbf{U}}_{i,j,k}^{n;LO}$ have been replaced by $\left(\bar{B}_{x;i+1/2,j,k}^{n;LO} + \bar{B}_{x;i-1/2,j,k}^{n;LO}\right)/2$, $\left(\bar{B}_{y;i,j+1/2,k}^{n;LO} + \bar{B}_{y;i,j-1/2,k}^{n;LO}\right)/2$ and $\left(\bar{B}_{z;i,j,k+1/2}^{n;LO} + \bar{B}_{z;i,j,k-1/2}^{n;LO}\right)/2$. We start by assuming that $\bar{\mathbf{U}}_{i,j,k}^{n;LO}$ is PCP. We then make the updates:-

$$\tilde{\mathbf{U}}_{i,j,k}^{n+1;LO} = \bar{\mathbf{U}}_{i,j,k}^{n;LO} - \frac{\Delta t}{\Delta x}\left(\bar{\mathbf{F}}_{i+1/2,j,k}^{LO} - \bar{\mathbf{F}}_{i-1/2,j,k}^{LO}\right) - \frac{\Delta t}{\Delta y}\left(\bar{\mathbf{G}}_{i,j+1/2,k}^{LO} - \bar{\mathbf{G}}_{i,j-1/2,k}^{LO}\right) - \frac{\Delta t}{\Delta z}\left(\bar{\mathbf{H}}_{i,j,k+1/2}^{LO} - \bar{\mathbf{H}}_{i,j,k-1/2}^{LO}\right) \quad (5.1)$$

$$\bar{B}_{x;i+1/2,j,k}^{n+1;LO} = \bar{B}_{x;i+1/2,j,k}^{n;LO} - \frac{\Delta t}{\Delta y \Delta z}\left(\Delta z \bar{E}_{z;i+1/2,j+1/2,k}^{LO} - \Delta z \bar{E}_{z;i+1/2,j-1/2,k}^{LO} + \Delta y \bar{E}_{y;i+1/2,j,k-1/2}^{LO} - \Delta y \bar{E}_{y;i+1/2,j,k+1/2}^{LO}\right)$$

(5.2)

$$\bar{B}_{y;i,j-1/2,k}^{n+1;LO} = \bar{B}_{y;i,j-1/2,k}^{n;LO} - \frac{\Delta t}{\Delta x \Delta z}\left(\Delta x \bar{E}_{x;i,j-1/2,k+1/2}^{LO} - \Delta x \bar{E}_{x;i,j-1/2,k-1/2}^{LO} + \Delta z \bar{E}_{z;i-1/2,j-1/2,k}^{LO} - \Delta z \bar{E}_{z;i+1/2,j-1/2,k}^{LO}\right)$$

(5.3)

$$\bar{B}_{z;i,j,k+1/2}^{n+1;LO} = \bar{B}_{z;i,j,k+1/2}^{n;LO} - \frac{\Delta t}{\Delta x \Delta y}\left(\Delta x \bar{E}_{x;i,j-1/2,k+1/2}^{LO} - \Delta x \bar{E}_{x;i,j+1/2,k+1/2}^{LO} + \Delta y \bar{E}_{y;i+1/2,j,k+1/2}^{LO} - \Delta y \bar{E}_{y;i-1/2,j,k+1/2}^{LO}\right)$$

(5.4)

Now realize that because $\bar{\mathbf{U}}_{i,j,k}^{n;LO}$ started off PCP, and because eqn. (5.1) is a first order update with a PCP-preserving flux, we are guaranteed that the updated zone-centered variable $\tilde{\mathbf{U}}_{i,j,k}^{n+1;LO}$ will be PCP. It is only when the sixth, seventh and eighth components of $\tilde{\mathbf{U}}_{i,j,k}^{n+1;LO}$ are replaced with $\left(\overline{B}_{x;i+1/2,j,k}^{n+1;LO} + \overline{B}_{x;i-1/2,j,k}^{n+1;LO}\right)/2$, $\left(\overline{B}_{y;i,j+1/2,k}^{n+1;LO} + \overline{B}_{y;i,j-1/2,k}^{n+1;LO}\right)/2$ and $\left(\overline{B}_{z;i,j,k+1/2}^{n+1;LO} + \overline{B}_{z;i,j,k-1/2}^{n+1;LO}\right)/2$ that we have the possibility of losing the PCP property in a few rare zones! After the replacement of the sixth, seventh and eighth components of $\tilde{\mathbf{U}}_{i,j,k}^{n+1;LO}$ by the facial averages, let us call the zone-centered variables $\overline{\mathbf{U}}_{i,j,k}^{n+1;LO}$. So it is this replacement by the facial averages that may, in a few rare zones, turn $\tilde{\mathbf{U}}_{i,j,k}^{n+1;LO}$ (which is PCP) into $\overline{\mathbf{U}}_{i,j,k}^{n+1;LO}$ (which may not be PCP). Our task is to fix this situation.

When the problem is cast this way, two clear fixes are obvious. The first fix is a fully conservative fix and we describe it in this paragraph. It finds its inspiration from recent work by Abgrall (2018, 2022). Realize that when $\overline{\mathbf{U}}_{i,j,k}^{n+1;LO}$ loses the PCP property, it does so because of the slightest discretization error; i.e. the facial fluxes that were used to update the zone-centered energy did not bring in enough energy from neighboring zones to keep $\overline{\mathbf{U}}_{i,j,k}^{n+1;LO}$ PCP. Let $[\mathbf{X}]_5$ denote the fifth component of a vector $\mathbf{X}$; this is just a choice of notation. We can ever so slightly modify the fifth component of the flux for the MHD case as follows:-

$$\left[\tilde{\mathbf{F}}_{i+1/2,j,k}^{LO}\right]_5 = \left[\overline{\mathbf{F}}_{i+1/2,j,k}^{LO}\right]_5 - \overline{w}_{i+1/2,j,k}^{-}\alpha \quad ; \quad \left[\tilde{\mathbf{F}}_{i-1/2,j,k}^{LO}\right]_5 = \left[\overline{\mathbf{F}}_{i-1/2,j,k}^{LO}\right]_5 + \overline{w}_{i-1/2,j,k}^{+}\alpha \quad ;$$
$$\left[\tilde{\mathbf{G}}_{i,j+1/2,k}^{LO}\right]_5 = \left[\overline{\mathbf{G}}_{i,j+1/2,k}^{LO}\right]_5 - \overline{w}_{i,j+1/2,k}^{-}\alpha \quad ; \quad \left[\tilde{\mathbf{G}}_{i,j-1/2,k}^{LO}\right]_5 = \left[\overline{\mathbf{G}}_{i,j-1/2,k}^{LO}\right]_5 + \overline{w}_{i,j-1/2,k}^{+}\alpha \quad ; \quad (5.5)$$
$$\left[\tilde{\mathbf{H}}_{i,j,k+1/2}^{LO}\right]_5 = \left[\overline{\mathbf{H}}_{i,j,k+1/2}^{LO}\right]_5 - \overline{w}_{i,j,k+1/2}^{-}\alpha \quad ; \quad \left[\tilde{\mathbf{H}}_{i,j,k-1/2}^{LO}\right]_5 = \left[\overline{\mathbf{H}}_{i,j,k-1/2}^{LO}\right]_5 + \overline{w}_{i,j,k-1/2}^{+}\alpha$$

The $\overline{w}$ variables will be described shortly and they are just a measure of a neighboring zone's ability to give enough energy to the troubled zone so that the troubled zone has a positive pressure. We can then quantify the variable "$\alpha$" in the above equation by setting it by using the equation:-

$$\frac{1}{8\pi}\left\{\begin{array}{l}\left(\left(\overline{B}_{x;i+1/2,j,k}^{n+1;LO} + \overline{B}_{x;i-1/2,j,k}^{n+1;LO}\right)/2\right)^2 + \left(\left(\overline{B}_{y;i,j+1/2,k}^{n+1;LO} + \overline{B}_{y;i,j-1/2,k}^{n+1;LO}\right)/2\right)^2 \\ + \left(\left(\overline{B}_{z;i,j,k+1/2}^{n+1;LO} + \overline{B}_{z;i,j,k-1/2}^{n+1;LO}\right)/2\right)^2 - \left(\left[\tilde{\mathbf{U}}_{i,j,k}^{n+1;LO}\right]_6\right)^2 - \left(\left[\tilde{\mathbf{U}}_{i,j,k}^{n+1;LO}\right]_7\right)^2 - \left(\left[\tilde{\mathbf{U}}_{i,j,k}^{n+1;LO}\right]_8\right)^2\end{array}\right\} = $$
$$\alpha\left\{\frac{\Delta t}{\Delta x}\left(\overline{w}_{i+1/2,j,k}^{-} + \overline{w}_{i-1/2,j,k}^{+}\right) + \frac{\Delta t}{\Delta y}\left(\overline{w}_{i,j+1/2,k}^{-} + \overline{w}_{i,j-1/2,k}^{+}\right) + \frac{\Delta t}{\Delta z}\left(\overline{w}_{i,j,k+1/2}^{-} + \overline{w}_{i,j,k-1/2}^{+}\right)\right\} \quad (5.6)$$

Then we can set

$$w^{-}_{i+1/2,j,k}=\left(\max\left(P^{n;LO}_{i+1,j,k}-P_{\min},0\right)\right)^{\chi}\ ;\ w^{+}_{i-1/2,j,k}=\left(\max\left(P^{n;LO}_{i-1,j,k}-P_{\min},0\right)\right)^{\chi}\ ;\ w^{-}_{i,j+1/2,k}=\left(\max\left(P^{n;LO}_{i,j+1,k}-P_{\min},0\right)\right)^{\chi}\ ;$$

$$w^{+}_{i,j-1/2,k}=\left(\max\left(P^{n;LO}_{i,j-1,k}-P_{\min},0\right)\right)^{\chi}\ ;\ w^{-}_{i,j,k+1/2}=\left(\max\left(P^{n;LO}_{i,j,k+1}-P_{\min},0\right)\right)^{\chi}\ ;\ w^{+}_{i,j,k-1/2}=\left(\max\left(P^{n;LO}_{i,j,k-1}-P_{\min},0\right)\right)^{\chi}$$

(5.7)

with $\chi \geq 1$. We use $\chi = 2$ and $P_{\min}$ is a user-settable small parameter. (We have set $P_{\min}=10^{-3}$ for all the tests reported here.) The pressures in the above equation stand as proxies for the thermal energy in those zones, but we can also use the actual thermal energy in those zones if we wish. The weights with overbars can then be normalized so that they add to unity in a WENO-style normalization that goes as follows

$$\overline{w}^{-}_{i+1/2,j,k} = w^{-}_{i+1/2,j,k}\Big/\left(w^{-}_{i+1/2,j,k}+w^{+}_{i-1/2,j,k}+w^{-}_{i,j+1/2,k}+w^{+}_{i,j-1/2,k}+w^{-}_{i,j,k+1/2}+w^{+}_{i,j,k-1/2}\right)$$

$$\overline{w}^{+}_{i-1/2,j,k} = w^{+}_{i-1/2,j,k}\Big/\left(w^{-}_{i+1/2,j,k}+w^{+}_{i-1/2,j,k}+w^{-}_{i,j+1/2,k}+w^{+}_{i,j-1/2,k}+w^{-}_{i,j,k+1/2}+w^{+}_{i,j,k-1/2}\right)$$

$$\overline{w}^{-}_{i,j+1/2,k} = w^{-}_{i,j+1/2,k}\Big/\left(w^{-}_{i+1/2,j,k}+w^{+}_{i-1/2,j,k}+w^{-}_{i,j+1/2,k}+w^{+}_{i,j-1/2,k}+w^{-}_{i,j,k+1/2}+w^{+}_{i,j,k-1/2}\right)$$

$$\overline{w}^{+}_{i,j-1/2,k} = w^{+}_{i,j-1/2,k}\Big/\left(w^{-}_{i+1/2,j,k}+w^{+}_{i-1/2,j,k}+w^{-}_{i,j+1/2,k}+w^{+}_{i,j-1/2,k}+w^{-}_{i,j,k+1/2}+w^{+}_{i,j,k-1/2}\right)$$

$$\overline{w}^{-}_{i,j,k+1/2} = w^{-}_{i,j,k+1/2}\Big/\left(w^{-}_{i+1/2,j,k}+w^{+}_{i-1/2,j,k}+w^{-}_{i,j+1/2,k}+w^{+}_{i,j-1/2,k}+w^{-}_{i,j,k+1/2}+w^{+}_{i,j,k-1/2}\right)$$

$$\overline{w}^{+}_{i,j,k-1/2} = w^{+}_{i,j,k-1/2}\Big/\left(w^{-}_{i+1/2,j,k}+w^{+}_{i-1/2,j,k}+w^{-}_{i,j+1/2,k}+w^{+}_{i,j-1/2,k}+w^{-}_{i,j,k+1/2}+w^{+}_{i,j,k-1/2}\right)$$

(5.8)

We see from this formulation that as long as a troubled zone has as at least one neighboring zone that can lend it some thermal energy, it will be able to obtain some extra energy to restore its positive pressure. This completes our description of the first fix, which is a fully conservative fix.

It is also possible that a zone may not have any von Neumann neighbors that have sufficient energy to lend it some thermal energy. Only in those rare occasions we take a different approach. Realize that $\tilde{\mathbf{U}}^{n+1;LO}_{i,j,k}$ is still PCP, so it still has positive pressure. Therefore, we already have a pressure in the troubled zone that is positive, except that it was obtained via the update in eqn. (5.1). We can reset the troubled zone to have that same positive pressure as follows:-

$$\left[\overline{\mathbf{U}}^{n+1;LO}_{i,j,k}\right]_5 = \left[\tilde{\mathbf{U}}^{n+1;LO}_{i,j,k}\right]_5 + \left[\mathbf{S}^{LO}_{i,j,k}\right]_5 \quad \text{with the definition}$$

$$\left[\mathbf{S}^{LO}_{i,j,k}\right]_5 \equiv \frac{1}{8\pi}\left\{\begin{array}{l}\left(\left(\overline{B}^{n+1;LO}_{x;i+1/2,j,k}+\overline{B}^{n+1;LO}_{x;i-1/2,j,k}\right)/2\right)^2 + \left(\left(\overline{B}^{n+1;LO}_{y;i,j+1/2,k}+\overline{B}^{n+1;LO}_{y;i,j-1/2,k}\right)/2\right)^2 \\ + \left(\left(\overline{B}^{n+1;LO}_{z;i,j,k+1/2}+\overline{B}^{n+1;LO}_{z;i,j,k-1/2}\right)/2\right)^2 - \left(\left[\tilde{\mathbf{U}}^{n+1;LO}_{i,j,k}\right]_6\right)^2 - \left(\left[\tilde{\mathbf{U}}^{n+1;LO}_{i,j,k}\right]_7\right)^2 - \left(\left[\tilde{\mathbf{U}}^{n+1;LO}_{i,j,k}\right]_8\right)^2\end{array}\right\}$$

(5.9)

The low order source term $\left[\mathbf{S}_{i,j,k}^{LO}\right]_5$ contributes only to the fifth component, which is the energy equation, for MHD. Only the 5th component of the source term vector $\mathbf{S}_{i,j,k}^{LO}$ is non-zero and that too only for zones that have lost the PCP property in the first order update. Note that eqn. (5.9) constitutes a very slight loss of energy conservation. But we will document this loss of energy conservation for very stringent problems in the results Section and show that this loss of energy conservation is absolutely inconsequential. The really big advantage of this lower order formulation is that we always have a PCP formulation that is divergence-free which can be used to guide the higher order scheme so that it always remains divergence-free and PCP. In the next Sub-Section we will show how we do this in the most unobtrusive of ways so that for the most part we only use the high order scheme, only resorting to the lower order scheme from this Sub-Section in zones where the PCP property may be lost. Here we have only described the fix in the context of non-relativistic MHD, but the extension to RMHD is described in the Appendix.

**5.2) Hybridizing the First Order PCP Scheme with the Higher Order Scheme**

In each zone $(i,j,k)$ we define a variable $\theta_{i,j,k}$. We will design a method such that when $\theta_{i,j,k}=1$ for all the zones, the scheme will exclusively be a high order scheme; this is the default when the astrophysical problem is not too stringent. When $\theta_{i,j,k}=0$ for all the zones, the scheme will exclusively be a first order scheme. Of course, it is not our intent to have $\theta_{i,j,k}=0$ in any of the zones, but for stringent problems, we might have $\theta_{i,j,k}=0$ in some of the zones. At each zone boundary we can define a flux by

$$\begin{aligned}
\bar{\mathbf{F}}_{i+1/2,j,k}^{\theta} &= \left(1-\theta_{i+1/2,j,k}^{face}\right)\bar{\mathbf{F}}_{i+1/2,j,k}^{LO} + \theta_{i+1/2,j,k}^{face}\bar{\mathbf{F}}_{i+1/2,j,k}^{HO} \quad \text{with} \quad \theta_{i+1/2,j,k}^{face} \equiv \min\left(\theta_{i,j,k},\theta_{i+1,j,k}\right);\\
\bar{\mathbf{G}}_{i,j+1/2,k}^{\theta} &= \left(1-\theta_{i,j+1/2,k}^{face}\right)\bar{\mathbf{G}}_{i,j+1/2,k}^{LO} + \theta_{i,j+1/2,k}^{face}\bar{\mathbf{G}}_{i,j+1/2,k}^{HO} \quad \text{with} \quad \theta_{i,j+1/2,k}^{face} \equiv \min\left(\theta_{i,j,k},\theta_{i,j+1,k}\right); \quad (5.10)\\
\bar{\mathbf{H}}_{i,j,k+1/2}^{\theta} &= \left(1-\theta_{i,j,k+1/2}^{face}\right)\bar{\mathbf{H}}_{i,j,k+1/2}^{LO} + \theta_{i,j,k+1/2}^{face}\bar{\mathbf{H}}_{i,j,k+1/2}^{HO} \quad \text{with} \quad \theta_{i,j,k+1/2}^{face} \equiv \min\left(\theta_{i,j,k},\theta_{i,j,k+1}\right)
\end{aligned}$$

The left panel of Fig. 6 shows how these facial values of $\theta$ are collocated. Likewise, at each edge we can define

$$\overline{E}^{\theta}_{z;i+1/2,j+1/2,k} = \left(1-\theta^{edge}_{i+1/2,j+1/2,k}\right)\overline{E}^{LO}_{z;i+1/2,j+1/2,k} + \theta^{edge}_{i+1/2,j+1/2,k}\overline{E}^{HO}_{z;i+1/2,j+1/2,k}$$

$$\text{with}\quad \theta^{edge}_{i+1/2,j+1/2,k} \equiv \min\left(\theta_{i,j,k},\theta_{i+1,j,k},\theta_{i,j+1,k},\theta_{i+1,j+1,k}\right);$$

$$\overline{E}^{\theta}_{y;i+1/2,j,k+1/2} = \left(1-\theta^{edge}_{i+1/2,j,k+1/2}\right)\overline{E}^{LO}_{y;i+1/2,j,k+1/2} + \theta^{edge}_{i+1/2,j,k+1/2}\overline{E}^{HO}_{y;i+1/2,j,k+1/2} \quad (5.11)$$

$$\text{with}\quad \theta^{edge}_{i+1/2,j,k+1/2} \equiv \min\left(\theta_{i,j,k},\theta_{i+1,j,k},\theta_{i,j,k+1},\theta_{i+1,j,k+1}\right);$$

$$\overline{E}^{\theta}_{x;i,j+1/2,k+1/2} = \left(1-\theta^{edge}_{i,j+1/2,k+1/2}\right)\overline{E}^{LO}_{x;i,j+1/2,k+1/2} + \theta^{edge}_{i,j+1/2,k+1/2}\overline{E}^{HO}_{x;i,j+1/2,k+1/2}$$

$$\text{with}\quad \theta^{edge}_{i,j+1/2,k+1/2} \equiv \min\left(\theta_{i,j,k},\theta_{i,j+1,k},\theta_{i,j,k+1},\theta_{i,j+1,k+1}\right)$$

The left panel of Fig. 6 shows how these edge values of $\theta$ are collocated. The right panel of Fig. 6 shows how these can be used to obtain electric fields at the edges. The above two equations show that the order of accuracy of the fluxes and electric fields can be locally lowered, as needed, to the point where the first order scheme always guarantees PCP behavior.

We now describe an iterative update strategy that does just that. We iterate over the whole mesh, starting the iteration process with $\theta_{i,j,k}=1$ for all the zones. This $\theta_{i,j,k}$ will be sequentially lowered for any zone that is troubled; but realize that it will only be lowered for the few zones that are troubled. For each iteration, eqns. (4.7) to (4.10) can be modified to become

$$\partial_t \overline{\mathbf{U}}^{\theta}_{i,j,k} = -\frac{1}{\Delta x}\left(\overline{\mathbf{F}}^{\theta}_{i+1/2,j,k} - \overline{\mathbf{F}}^{\theta}_{i-1/2,j,k}\right) - \frac{1}{\Delta y}\left(\overline{\mathbf{G}}^{\theta}_{i,j+1/2,k} - \overline{\mathbf{G}}^{\theta}_{i,j-1/2,k}\right) - \frac{1}{\Delta z}\left(\overline{\mathbf{H}}^{\theta}_{i,j,k+1/2} - \overline{\mathbf{H}}^{\theta}_{i,j,k-1/2}\right)$$
$$+ \frac{1}{\Delta t}\left(1-\theta_{i,j,k}\right)\mathbf{S}^{LO}_{i,j,k} \quad (5.12)$$

$$\partial_t \overline{B}^{\theta}_{x;i+1/2,j,k} = -\frac{1}{\Delta y \Delta z}\left(\Delta z \overline{E}^{\theta}_{z;i+1/2,j+1/2,k} - \Delta z \overline{E}^{\theta}_{z;i+1/2,j-1/2,k} + \Delta y \overline{E}^{\theta}_{y;i+1/2,j,k-1/2} - \Delta y \overline{E}^{\theta}_{y;i+1/2,j,k+1/2}\right) \quad (5.13)$$

$$\partial_t \overline{B}^{\theta}_{y;i,j-1/2,k} = -\frac{1}{\Delta x \Delta z}\left(\Delta x \overline{E}^{\theta}_{x;i,j-1/2,k+1/2} - \Delta x \overline{E}^{\theta}_{x;i,j-1/2,k-1/2} + \Delta z \overline{E}^{\theta}_{z;i-1/2,j-1/2,k} - \Delta z \overline{E}^{\theta}_{z;i+1/2,j-1/2,k}\right) \quad (5.14)$$

$$\partial_t \overline{B}^{\theta}_{z;i,j,k+1/2} = -\frac{1}{\Delta x \Delta y}\left(\Delta x \overline{E}^{\theta}_{x;i,j-1/2,k+1/2} - \Delta x \overline{E}^{\theta}_{x;i,j+1/2,k+1/2} + \Delta y \overline{E}^{\theta}_{y;i+1/2,j,k+1/2} - \Delta y \overline{E}^{\theta}_{y;i-1/2,j,k+1/2}\right) \quad (5.15)$$

Fig. 7 shows us schematically how our update strategy allows us to access any accuracy of update for the electric fields, going from the highest order accuracy to the lowest order accuracy. In this figure and the text that follows, the high order components are superscripted with "HO"; and low order components are superscripted with "LO". Using these, the forward Euler approximants can be constructed in eqns. (4.11) to (4.13). If a zone in those forward Euler approximants is not within the PCP domain, its local $\theta_{i,j,k}$ will be lowered even further in each successive iteration till it becomes PCP. Realize that this procedure is fully explicit; i.e. it does not require any fluxes or electric fields that are obtained through an implicit process. Realize too that this process is guaranteed to ensure that all zones are in the PCP domain.

**6) Step by Step Description of the PCP Method**

For the sake of simplicity, we describe the implementation in terms of the LLF Riemann solver. The method is implemented using the following steps:-

**Step 1)** The magnetic field components are reconstructed within the faces and the ADP algorithm from Balsara, Samantaray and Subramanian (2023) is used to obtain a higher order reconstruction of the entire magnetic field within the entire volume each zone. This also enables us to evaluate the magnetic field at each zone center. The 6$^{th}$, 7$^{th}$ and 8$^{th}$ components of $\overline{\mathbf{U}}_{i,j,k}$ are reset using the zone-averaged magnetic fields that were evaluated in this step.

**Step 2)** This is the transcription step. It is important because we make a sequence of transcriptions of the variables, going from $\overline{\mathbf{U}}_{i,j,k}$ to $\mathbf{U}_{i,j,k}$ to $\mathbf{V}_{i,j,k}$ to $\overline{\mathbf{V}}_{i,j,k}$. At each step in the transcription we have to check that the states remain in the PCP domain. Using eqn. (4.1) or eqn. (4.3), we go from zone-averaged values for the conserved variables $\overline{\mathbf{U}}_{i,j,k}$ to zone-centered point values for the conserved variables $\mathbf{U}_{i,j,k}$. (Note too that if a higher order transcription from zone-averaged $\overline{\mathbf{U}}_{i,j,k}$ to point value $\mathbf{U}_{i,j,k}$ produces a $\mathbf{U}_{i,j,k}$ that is not PCP, then we always check for this possibility and give ourselves the option to use a lower order transcription.) From $\mathbf{U}_{i,j,k}$, we can evaluate the vector of primitive variables $\mathbf{V}_{i,j,k}$ as zone-centered point values. From $\mathbf{V}_{i,j,k}$, we use (4.2) and (4.4) to build the zone-averaged values for the primitive variables $\overline{\mathbf{V}}_{i,j,k}$. (Because we check at each stage in this transcription, $\mathbf{V}_{i,j,k}$ and $\overline{\mathbf{V}}_{i,j,k}$ are also kept within the PCP domain.) The point values of the primitive variables $\mathbf{V}_{i,j,k}$ are also used to build the zone-centered point values for the electric field vector $\mathbf{E}_{i,j,k}$.

**Step 3)** The zone-centered point values for the electric field vector $\mathbf{E}_{i,j,k}$ are then *interpolated* with high order using WENO interpolation with the appropriate order of accuracy. The zone-averaged $\overline{\mathbf{V}}_{i,j,k}$ are *reconstructed* using WENO reconstruction with the appropriate order of accuracy. If needed, we use eqns. (4.5) and (4.6) to bring the higher order reconstructed primitive functions, $\hat{\mathbf{V}}_{i,j,k}(x,y,z)$, within the PCP domain. This ensures that Riemann solvers will always get physically realizable values.

**Step 4)** At each zone boundary, use the reconstructed primitive variables $\hat{\mathbf{V}}_{i,j,k}(x,y,z)$ to build the extremal wave speeds at that zone boundary as well as the high order numerical fluxes at that zone boundary. In fact, the extremal wave speeds at a zone boundary are a natural corollary to invoking the Riemann solvers at that zone boundary in order to get the numerical fluxes. For example, at each x-boundary $(i+1/2, j, k)$ we build $S_{R;i+1/2,j,k}$ and $S_{L;i+1/2,j,k}$ as well as the higher order x-flux $\overline{\mathbf{F}}^{HO}_{i+1/2,j,k}$. This also ensures that we can find extremal wave speeds at each edge. For example, for the z-edge $(i+1/2, j+1/2, k)$ we can find $S_{R;i+1/2,j+1/2,k} = \max\left(S_{R;i+1/2,j,k}, S_{R;i+1/2,j+1,k}\right)$, $S_{L;i+1/2,j+1/2,k} = \min\left(S_{L;i+1/2,j,k}, S_{L;i+1/2,j+1,k}\right)$, $S_{U;i+1/2,j+1/2,k} = \max\left(S_{U;i,j+1/2,k}, S_{U;i+1,j+1/2,k}\right)$ and $S_{D;i+1/2,j+1/2,k} = \min\left(S_{D;i,j+1/2,k}, S_{D;i+1,j+1/2,k}\right)$ which will be useful for the construction of the z-component of the electric field at that edge.

**Step 5)** Now that we have the extremal wave speeds (in both the transverse directions) at each edge, we can use the facially reconstructed magnetic field components from Step 1, as well as the interpolated electric fields from Step 3 to build higher order numerical electric fields at that edge. For example, from eqn. (3.9), we can find a high order z-component of electric field, $\overline{E}^{HO}_{z;i+1/2,j+1/2,k}$, at the z-edge $(i+1/2, j+1/2, k)$. (Realize that if the only thing that we desire is a higher order scheme then eqns. (4.7) to (4.10), along with the update equations (4.11) or (4.12) or (4.13), are fully adequate for giving us a very competent higher order scheme.)

**Step 6)** If PCP methods are needed, it is important to realize that at each fractional step we will additionally need to run an inexpensive low order scheme. The philosophy of PCP is that we have one lower order scheme that is guaranteed to be PCP, which guides the higher order scheme so that it remains PCP. This means that at each fractional step we will also send in the same data to the lower order scheme that we use for the higher order scheme.

**Step 7)** If PCP methods are needed, obtain the low order fluxes and electric fields that are described in Section 5.1 by using the low order PCP scheme. This is done once for each Runge-Kutta stage in the time update.

**Step 8)** If PCP methods are needed, in each zone we start with a variable $\theta_{i,j,k} = 1$. We then begin a sequence of iterations over the entire mesh. Zones that are within the PCP domain are left alone within each iteration. Zones whose update takes them outside the PCP domain are given a lower value of $\theta_{i,j,k}$, as long as we retain $\theta_{i,j,k} \geq 0$. In this fashion, we let the numerical scheme itself decide where it needs to use elements of a lower order scheme and how much of the lower order scheme is needed to render the entire solution within the PCP domain. This is done for each Runge-Kutta stage in the time update. (In this paper we have shown only the most stringent MHD and RMHD problems that are known in the literature. In all cases, we were able to bring the solution within the PCP domain within 10 to 20 iterations.)

## 7) Accuracy Analysis and Dissipation

### 7.1) Accuracy Analysis for MHD and RMHD

In this Sub-section we consider the two-dimensional Magnetohydrodynamics (MHD) and Relativistic-Magnetohydrodynamics (RMHD) systems from Section 2 and study the accuracy of the presented schemes for these systems.

For the MHD system, we consider the two-dimensional vortex problem from Balsara (2004). The problem consists of a smoothly varying and dynamically stable vortex that moves diagonally in a periodic domain. The explicit expressions for the setup have been provided in Balsara (2004). Therefore, we do not describe the setup details here. There is, however, one point of difference. In Balsara (2004) we recommended doing this problem on a periodic domain ranging over $[-5,5] \times [-5,5]$ and running to a final time of 10 units. That choice of domain is fine for second order schemes. But it is important to realize that a smooth idealized vortex has, in principle, an infinite extent. To carry out an accuracy analysis, we have to truncate the problem and use periodic boundaries. For lower order schemes, this truncation of the vortex at the periodic boundaries does not have a substantial impact on the accuracy. But higher order schemes are very sensitive and can pick up differences from the boundary conditions if the boundary is not set far enough away. As a result, we now recommend doing this problem on a periodic domain that ranges over $[-10,10] \times [-10,10]$ and running to a final time of 20 units. In Fig. 8 we show the $L_1$ and $L_\infty$ errors for schemes ranging from second order accurate to sixth order accurate. The slopes in the left upper corner of this figure also show the theoretical accuracies. The top row of Fig. 8 shows the error in the x-momentum and the x-magnetic field for non-relativistic MHD. We observe that the schemes we have presented are able to reach their design accuracy. Indeed, we see from Fig. 8 that the fourth and fifth order schemes are quite close because such problems sometimes show serendipitous increase in order. So the fourth order scheme has almost the same accuracy as the fifth order scheme. But it is useful to realize that this serendipitous increase in order is problem-specific.

For the RMHD system, Balsara and Kim (2016) have presented a two-dimensional vortex problem with detailed derivation and setup expressions. Because they were displaying some fourth order schemes, they already recommended using a domain that ranges over $[-10,10] \times [-10,10]$. That set-up is intrinsically based on using a larger domain. We consider the same setup here to study the accuracy of the presented schemes. The lower row of Fig. 8 shows the error in the x-momentum and the x-magnetic field for the RMHD vortex. Observe that the schemes we present here are able to reach their design accuracy for the RMHD system at all the orders in the $L_1$ and $L_\infty$ norms. For the $L_1$ and $L_\infty$ errors in the $B_x$ variable for the RMHD vortex, the kinks in the red and purple curves show that the higher order accuracy is asymptotically reached only at the highest resolutions. This is a well-known fact for all higher order schemes:- The theory for hyperbolic system scheme design only says that the schemes reach their design accuracies in the $L_1$ norm and that too at the highest resolutions (Harten 1983). As one can see from the last two panels in the lower row of Fig. 8, one might have to go to rather high resolutions before the design accuracy of the scheme is achieved. In the RMHD vortex we find that the fifth order scheme is substantially more accurate compared to the fourth order scheme. This illustrates that the accuracy improvement that we found for the MHD vortex is indeed serendipitous.

**7.2) Controlling Dissipation**

For numerous astrophysical problems it is very useful to control the dissipation. This is especially so for the magnetic field in such areas as turbulence simulations and dynamo studies. Appendix B explains why the dissipation shows a strong dependence on the quality of the Riemann solver, especially when a lower order scheme is used. Leidi *et al.* (2022) have repurposed the magnetized vortex problem from Balsara (2004) and used it as a way of showing the dissipation of the magnetic field. If the magnetic field in the magnetized vortex retains most of its strength at the end of one periodic propagation, then that is taken as an indication that dissipation is properly controlled by the scheme. The top row of Fig. 9 shows the magnetic energy after one complete propagation across the domain when the multidimensional LLF Riemann solver is used. The bottom row of Fig. 9 shows the magnetic energy after one complete propagation across the domain when the multidimensional HLL Riemann solver is used. In Fig. 9, we normalize the obtained magnetic energy by the maximum initial magnetic energy. At low orders, say second and third order, there is a substantial difference between the magnetic energy that is retained on the mesh when the LLF Riemann solver is used compared to when the HLL Riemann solver is used. Therefore, at lower orders, the quality of the multidimensional Riemann solver does substantially affect the retention of magnetic energy on the mesh. Indeed, Balsara and Nkonga (2017) have even presented a multidimensional Riemann solver that retains all the contributions from Alfven wave propagation (and indeed from all the MHD waves) in a truly multidimensional sense. However, we also see from Fig. 9 that when one goes to fourth and higher orders, the difference between the multidimensional LLF Riemann solver and the multidimensional HLL Riemann solver is absolutely minuscule. This is because the higher order reconstruction does an excellent job of minimizing the jumps in the states that are fed into the Riemann solver. Minimizing the jumps in the magnetic field components also minimizes the dissipation. This might be a new insight that might not have been appreciated in the MHD literature.

## 8) Stringent Test Problems

We present several two-dimensional stringent test problems for the MHD and RMHD systems. For each of the problems, we take a CFL number of 0.4 unless stated otherwise.

### 8.1) Stringent Blast-Wave problem for the MHD system

In this Sub-Section, we consider the blast-wave problem for the MHD system. The blast problem with a moderate magnetized magnetic field is described in Balsara & Spicer (1999), and the strongly magnetized blast problem is described in Wu & Shu (2018). Here, we only consider the strongly magnetized, very stringent test problem from Wu & Shu (2018) because it corresponds to a very low value of plasma beta $(\beta = 2.51 \times 10^{-6})$. The problem was set up on a two-dimensional domain that spans $[-0.6, 0.6] \times [-0.6, 0.6]$. We use the same setup as described in Wu & Shu (2018). Initially, the density is uniformly set to unity and the velocity is set to zero. The pressure is uniformly set to 0.1 except within a central circle of radius 0.1, where it is elevated to 10000. Additionally, a magnetic field with a magnitude of 1000 is initialized along the *x*-direction that makes the problem very challenging for numerical schemes. We simulate the problem on a 2D grid consisting of 400×400 zones until a final time of *t* = 0.001. We use the fourth-order scheme and show results for the density profile, pressure profile, magnitude-squared of the velocity vector, and magnitude-squared of the magnetic field vector in Figs. 10a to 10d respectively. Fig 10e also shows the value of the $\theta$-variable at the final time. For this problem, there is no energy fed in from the boundaries, so any energy violation can only come from the source term in eqn. (5.9). Therefore, the total energy in the simulation as a function of time is also shown in Fig. 10e. Notice that the original magnetic field is so strong in this problem that the magnetic pressure at late times has only a 17.5% variation in Fig. 10d despite the intense blast wave that is set off. The results obtained are consistent with those reported in Wu & Shu (2018), highlighting the performance of the presented schemes.

## 8.2) Stringent Astrophysical jet problem for the MHD system

We now consider the astrophysical jet problem (with Mach number of 800) from Balsara (2012). Following Wu & Shu (2018), a magnetic field is added to this problem to simulate the MHD jet flows. The presence of a magnetic field makes this test problem even more extreme. We only consider the extremely strong magnetized case, where we have $B_a = \sqrt{20000}$ (corresponding $\beta_a = 10^{-4}$). The problem was set up on a two-dimensional domain that spans $[-0.5, 0.5] \times [0.0, 1.5]$. We refer the readers to Wu & Shu (2018) for the detailed setup of the problem. We run the test case on a 2D grid consisting of 400×600 zones until a stopping time of *t*=0.002. A sixth-order accurate scheme was used for the run. Figures 11a and 11b show the resulting density and pressure profiles on logarithmic scales; Fig. 11c shows the square of the magnitude of the magnetic field vector; Fig. 11d shows the $\theta$-variable at the final time. We observe that the working surface of the jet and the cocoon are well captured for this extreme test; hence showing the robust performance of the proposed methods.

**8.3) Stringent Blast-Wave problem for the RMHD system**

In this Sub-Section, we consider the blast problem for the RMHD system. The non-relativistic version of this test problem was first presented in Balsara and Spicer (1999) and was extended to RMHD system by Komissarov (1999). Several extreme variants of this test problem have been presented in Wu & Shu (2021). We focus on one of the variants presented in Wu & Shu (2021), characterized by a plasma beta $\beta = 2.5 \times 10^{-6}$. The test problem is set up on a two-dimensional square domain that spans $[-6,6] \times [-6,6]$. Within a radius of 0.8, the explosion zone has a density of $10^{-2}$ and a pressure of 1. Outside a radius of 1 unit, the ambient medium has a density of $10^{-4}$ and a pressure of $5 \times 10^{-4}$. A linear taper is applied to the density and pressure from a radius of 0.8 to 1. Accordingly, both the density and pressure linearly decrease with increasing radius in that range of radii. The magnetic field is initialized in the x-direction and has a magnitude of 20. The corresponding plasma beta is very low $(\beta = 2.5 \times 10^{-6})$, which makes the test case stringent. The polytropic index of $\Gamma = 4/3$ is used in this problem. The simulation is run to a final time of 4 using a fifth-order accurate scheme on a grid consisting of $400 \times 400$ zones. The results have been presented in Fig. 12. Figure 12a shows the resulting density profile, Fig. 12b shows the pressure profile, and Fig. 12c shows the normed-squared value of the variable $\gamma \mathbf{v}$, and Fig. 12d shows the magnitude-squared of the magnetic field at time $t=4$. Fig. 12e shows the value of the $\theta$-variable at the final time, and Fig. 12f shows the total energy in the simulation as a function of time. Notice that the original magnetic field is so strong in this problem that the magnetic pressure at late times has less than 1% variation in Fig. 12d. We see a close resemblance between the results we obtained and the results presented in Wu & Shu (2021).

**8.4) Stringent Astrophysical jet problem for the RMHD system**

In this Sub-Section, we focus on the supersonic astrophysical jet problem from Wu & Shu (2021) for the RMHD system. The problem is initialized on a two-dimensional domain that spans $[-12,12] \times [0,25]$. The domain is filled with a uniform medium, where the density, velocity, and pressure are given by $\rho = 1$, $\mathbf{v} = \mathbf{0}$, and $p = 2.3536241 \times 10^{-5}$, respectively. Along the y-direction, a magnetic field $\mathbf{B} = \left(0, \sqrt{2000p}, 0\right)$ is initialized. The corresponding initial Lorentz factor is $\gamma \approx 7.09$ and the corresponding relativistic Mach number is $M_r \approx 354.37$. The very high Mach number and substantial Lorentz factor make the simulation of this problem highly challenging. At the bottom boundary $(y = 0)$, a dense jet is injected through the inlet part $|x| < 0.5$ with the states $\left(\rho, v_x, v_y, v_z, p, B_x, B_y, B_z\right) = \left(0.1, 0, 0.99, 0, 2.3536241 \times 10^{-5}, 0, \sqrt{2000p}, 0\right)$. Outflow boundary conditions are employed at all the other boundaries. The adiabatic constant is set as $\Gamma = 5/3$. The corresponding plasma beta is $\beta = 10^{-3}$, which makes the simulation even more challenging. We run the test case on a 2D grid consisting of 720×750 zones until a stopping time of $t=30$. A fourth-order accurate scheme was used for the run. Figures 13a and 13b show the resulting density and pressure profiles on logarithmic scales; Fig. 13c shows the square of the magnitude of the magnetic field vector, Fig. 13d shows the $\theta$-variable at the final time. We see that the Mach shock wave and interfaces are well captured for this extreme case and the results closely match the results presented in Wu & Shu (2021). This demonstrates the robust performance of the proposed methods.

**8.5) Energy Conservation**

From the source term in eqn. (5.9) we see that the energy may indeed not be exactly conserved when a zone that is updated with a higher order scheme takes a correction from the first order code. In such circumstances, we want the energy correction to be minimal. We would also like to demonstrate that when the same problem is solved on meshes with increasing resolution, the higher resolution meshes take smaller amounts of energy correction. For jet problems, there is energy injection from the base of the jet. This makes the jet problems unsuitable for demonstrating energy conservation. However, for both the blast test problems, there is no energy input from the boundaries. As a result, we should be able to evaluate energy conservation for such problems. We solve the same problems that were described in Sub-sections 8.1 and 8.3, but this time we solve them on a domain that is four times larger and run them for a time that is also four times longer. Thus we have a sequence of meshes with 400×400 zones, 800×800 zones and 1600×1600 zones. Fig. 10f shows the percent energy violation for the non-relativistic blast wave problem on this sequence of meshes. We see that the loss of energy conservation is of the order of 0.005% on the coarsest mesh and decreases as the mesh becomes more refined. Fig. 12f shows the percent energy violation for the relativistic blast wave problem on this sequence of meshes. We see that the loss of energy conservation is of the order of 0.00008% on the coarsest mesh and decreases as the mesh becomes more refined. This shows that even for the most stringent of problems, the energy tends towards near perfect conservation as the resolution of the problem is increased. We also recall, that if the problem is not extremely stringent, i.e. if there are no troubled zones, then the energy conservation is exact.

**9) Conclusions**

While second order schemes have been the mainstay of computational astrophysics in the past, there is an emerging drive to carry out astrophysical simulations with very high order of accuracy. All of these schemes, including of course the second order ones, can produce unphysical flow variables. This almost inevitably happens in problems with large Mach numbers, unusually large Lorentz factors and very strong magnetic fields. Code crashes, and a concomitant sense of frustration, are the inevitable result when that happens. The present paper is designed to rectify this situation.

In this paper an extremely inexpensive form of multidimensional Riemann solver is presented in Section 3. (Appendix B shows why the multidimensional Riemann solver is needed in MHD using a physically-motivated explanation.) In Section 4 we show the novel result that very high order reconstruction can be done in the primitive variables. In that Section, we also show that this reconstruction can always be modified to ensure that the values fed to a Riemann solver are always within the PCP domain. This eliminates one source of failure, i.e. situations where Riemann solvers are fed unphysical values resulting in a code crash. However, it does not ensure that the time update that results will keep the solution within the PCP domain.

We present a very lightweight first order scheme in Section 5.1 that will always keep the solution within the PCP domain. In Section 5.2 we show how this first order scheme can be nonlinearly hybridized with the higher order scheme to ensure that the overall time update is in the PCP domain. This nonlinear hybridization is also lightweight. Section 6 provides a point by point implementation strategy. Section 7 shows that the method is perfectly unobtrusive for higher order simulations that do not violate the PCP condition. Section 8 shows the results of some of the most stringent test problems – i.e. the ones that are known in the literature to cause code crashes. We show that our method successfully overcomes all those situations while running with a robust CFL number.

**Acknowledgements**

DSB acknowledges support via NSF grants NSF-AST-2434532 and NASA grant 80NSSC22K0628. RK acknowledges support via SNSF grant 10001228. We also acknowledge a travel grant for CS by Notre Dame International that made this collaboration possible.

**Appendix A: Positivity fix for the RMHD system**

In Sub-Section 5.1 we described a positivity fix for non-relativistic magnetohydrodynamics. Here, we describe a positivity fix for the relativistic magnetohydrodynamics (RMHD) system. Let us recall some mathematical terms from Sub-Section 5.1. We have $\bar{\mathbf{U}}_{i,j,k}^{n;LO}$ where the sixth, seventh and eighth components of $\bar{\mathbf{U}}_{i,j,k}^{n;LO}$ replaced by $\left(\bar{B}_{x;i+1/2,j,k}^{n;LO} + \bar{B}_{x;i-1/2,j,k}^{n;LO}\right)/2$, $\left(\bar{B}_{y;i,j+1/2,k}^{n;LO} + \bar{B}_{y;i,j-1/2,k}^{n;LO}\right)/2$ and $\left(\bar{B}_{z;i,j,k+1/2}^{n;LO} + \bar{B}_{z;i,j,k-1/2}^{n;LO}\right)/2$. We start by assuming that $\bar{\mathbf{U}}_{i,j,k}^{n;LO}$ is PCP. Similar to Sub-Section 5.1, we then make the update of this first order scheme by using eqns. (5.1) to (5.4). Now realize that because $\bar{\mathbf{U}}_{i,j,k}^{n;LO}$ started off PCP, and because eqn. (5.1) is a first order update with a PCP-preserving flux, we are guaranteed that the updated zone-centered variable $\tilde{\mathbf{U}}_{i,j,k}^{n+1;LO}$ will be PCP. It is only when the sixth, seventh and eighth components of $\tilde{\mathbf{U}}_{i,j,k}^{n+1;LO}$ are replaced with $\left(\bar{B}_{x;i+1/2,j,k}^{n+1;LO} + \bar{B}_{x;i-1/2,j,k}^{n+1;LO}\right)/2$, $\left(\bar{B}_{y;i,j+1/2,k}^{n+1;LO} + \bar{B}_{y;i,j-1/2,k}^{n+1;LO}\right)/2$ and $\left(\bar{B}_{z;i,j,k+1/2}^{n+1;LO} + \bar{B}_{z;i,j,k-1/2}^{n+1;LO}\right)/2$ that we have the possibility of losing the PCP property in a few rare zones! After the replacement of the sixth, seventh and eighth components of $\tilde{\mathbf{U}}_{i,j,k}^{n+1;LO}$ by the facial averages, let us call the zone-centered variables $\bar{\mathbf{U}}_{i,j,k}^{n+1;LO}$. So it is this replacement by the facial averages that may, in rare occasions, turn $\tilde{\mathbf{U}}_{i,j,k}^{n+1;LO}$ (which is PCP) into $\bar{\mathbf{U}}_{i,j,k}^{n+1;LO}$ (which may not be PCP in a few rare zones). Our task is to fix this situation.

Realize that $\tilde{\mathbf{U}}_{i,j,k}^{n+1;LO}$ is PCP, therefore it has positive density, positive pressure and bounded velocity vector $\left(|\tilde{\mathbf{v}}|<1\right)$. We denote the corresponding primitives by $\left[\tilde{\rho},\tilde{\mathbf{v}},\tilde{p}\right]_{i,j,k}$. Therefore, we already have primitives $\left[\tilde{\rho},\tilde{\mathbf{v}},\tilde{p}\right]_{i,j,k}$ in the troubled zone that are PCP, except that they were obtained via the update in eqn. (5.1). We define a new 8 component vector $\widehat{\mathbf{V}}_{i,j,k}^{n+1;LO}$ that has first 5 components same as the primitives $\left[\tilde{\rho},\tilde{\mathbf{v}},\tilde{p}\right]_{i,j,k}$. The sixth, seventh and eighth components of $\widehat{\mathbf{V}}_{i,j,k}^{n+1;LO}$ are assigned the values $\left(\bar{B}_{x;i+1/2,j,k}^{n+1;LO} + \bar{B}_{x;i-1/2,j,k}^{n+1;LO}\right)/2$, $\left(\bar{B}_{y;i,j+1/2,k}^{n+1;LO} + \bar{B}_{y;i,j-1/2,k}^{n+1;LO}\right)/2$ and $\left(\bar{B}_{z;i,j,k+1/2}^{n+1;LO} + \bar{B}_{z;i,j,k-1/2}^{n+1;LO}\right)/2$ respectively. From the primitives $\widehat{\mathbf{V}}_{i,j,k}^{n+1;LO}$ we obtain the corresponding conservative variables in $\widehat{\mathbf{U}}_{i,j,k}^{n+1;LO}$. Realize that, just by the definition, $\widehat{\mathbf{U}}_{i,j,k}^{n+1;LO}$ is PCP. Therefore, to obtain the PCP states in a troubled zone, we reset the states in the troubled zone as follows:-

$$\left[\bar{\mathbf{U}}_{i,j,k}^{n+1;LO}\right]_m = \left[\widehat{\mathbf{U}}_{i,j,k}^{n+1;LO}\right]_m \quad \text{for } m = 1,2,...,5. \tag{A.5}$$

This completes the description of a low order PCP strategy for the RMHD system. It is integrated into the higher order scheme as described in the text.

**Appendix B: Why do 1D and 2D Riemann Solvers Work?**

Since this is an astrophysics paper, it is possible to give the reader a physics-based argument as to why 1D and 2D Riemann solvers work. We can also explain the physics-driven connection between 1D and 2D Riemann solvers. To see that physics-motivated perspective, we will focus on the LLF Riemann solver. Consider the x-velocity equation for 1D isothermal flow. To keep the discussion very simple, say that the flow even has constant density. Then the evolutionary equation with a viscosity "$\mu$" is given by:

$$\frac{\partial v_x}{\partial t} + \frac{\partial}{\partial x}\left(v_x^2 - \mu\frac{\partial v_x}{\partial x}\right) = 0.$$

The form of the viscosity term in the above equation is the only tensorially invariant form that would guarantee conservation while allowing a dissipative term to operate. Since the above equation has a parabolic term, it is always expected to have a smooth solution on a fine mesh where the viscous terms are well-resolved. But we want the solution to be free of oscillations on coarser meshes even when the viscous terms are not well-resolved. This requires enhancing the dissipation term. The above equation will be numerically stabilized on a mesh with zones of any size if we use a numerical flux that is given by

$$F_{LLF}^* = \frac{1}{2}\left(v_{xR}^2 + v_{xL}^2\right) - \mu\left(v_{xR} - v_{xL}\right).$$

The genesis of the above LLF flux stems from the realization that with $\mu \to S/2$ we will get this desired stabilization on any mesh as long as "$S$" is the largest local signal speed in the problem! In that limit, we can clearly see the centered part of the numerical flux as well as the dissipation terms that are needed for its numerical stabilization. (Of course, if one is working on a fine mesh that is fine enough to capture viscous scales, one would want "$\eta$" to be the smaller value between the physical viscosity and the numerically motivated one.) This preamble enables us to understand the stabilization in the 2D LLF Riemann solver that is described in the next paragraph.

The induction equation for resistive MHD with a resistivity "$\eta$" can now be written as:

$$\frac{\partial \mathbf{B}}{\partial t} + \nabla \times \left(-\mathbf{v} \times \mathbf{B} + \eta \nabla \times \mathbf{B}\right) = 0.$$

The form of the resistivity term in the above equation is the only tensorially invariant form that would guarantee divergence-free evolution while allowing a dissipative term to operate. Since the above equation also has a parabolic term ($\nabla \times (\eta \nabla \times \mathbf{B}) = -\eta \nabla^2 \mathbf{B}$), it is always expected to have a smooth solution on a fine mesh where the resistive terms are well-resolved. But we want the solution to be free of oscillations on coarser meshes even when the resistive terms are not well-resolved. This requires enhancing the dissipation term. The above equation will be numerically stabilized on a mesh with zones of any size if we use a numerical electric field that is given by

$$E_z^{**} = \left(E_{zRU} + E_{zLU} + E_{zLD} + E_{zRD}\right)/4 + \eta\left[B_{xD} - B_{xU} + B_{yR} - B_{yL}\right].$$

By setting $E_{zRU} \to (-\mathbf{v} \times \mathbf{B})_{zRU}$, $E_{zLU} \to (-\mathbf{v} \times \mathbf{B})_{zLU}$, $E_{zLD} \to (-\mathbf{v} \times \mathbf{B})_{zLD}$ and $E_{zRD} \to (-\mathbf{v} \times \mathbf{B})_{zRD}$ we can clearly identify the centered part of the numerical electric field that is to be used; please see Fig. 2. We can also see that the solution is stabilized on any mesh if $\eta \to S/2$. A von Neumann stability analysis of the multidimensional induction equation by Balsara and Käppeli (2017) shows that $\eta \to S/2$ is indeed the right setting for numerical stability. It is easy to see that the discrete part of the curl is approximated as $\nabla \times \mathbf{B} \to \left[ B_{xD} - B_{xU} + B_{yR} - B_{yL} \right]$; please see Fig. 2. (Of course, if one is working on a fine mesh that is fine enough to capture resistive scales, one would want "$\eta$" to be the smaller value between the physical resistivity and the numerically motivated one.)

In production codes, one goes through many more details to produce much better Riemann solvers with more desirable properties. The multidimensional HLL Riemann solver presented in this paper is a case in point because it permits the multidimensional upwinding to be maintained. A better Riemann solver provides a larger advantage to a numerical scheme especially when the order of accuracy of the numerical scheme is low. As the order of accuracy of the reconstruction improves, the jumps in the reconstructed variables that contribute to the dissipation terms become progressively smaller (as long as the problem is appropriately smooth). Therefore, very high order schemes can usually operate without much increase in the numerical dissipation even when a lower grade Riemann solver is used. This trend is clearly visible in Fig. 9 of this paper.

**Figure Captions**

*Fig. 1 Most astrophysics and space physics use the Yee discretization shown here. The figure shows that the primal variables of the magnetic field, given by the normal components of the magnetic field, are facially-collocated. The components of the primal magnetic field vector are shown by the thick arrows. The overbars on the magnetic field components indicate that these are facially averaged. They undergo an update from the induction equation. The edge-collocated electric fields, which are used for updating the facial magnetic induction components, are shown by the thin arrows close to the appropriate edge. They too have overbars to indicate that they are edge-averaged. The superscript "num" for the electric field components indicates that they are multidimensionally stabilized and, therefore, suitable for use in a numerical scheme.*

*Fig. 2 shows four zones in the xy-plane that come together at the z-edge of a three-dimensional mesh. Since the mesh is viewed from the top in plan view, the z-edge is shown by the black dot and the four abutting zones are shown as four squares. The four incoming states have subscripts given by "RU" for right-upper; "LU" for left-upper; "LD" for left-down and "RD" for right-down. Fig 2 shows the situation before the states start interacting via four one-dimensional and one multidimensional Riemann problems. The thin oblique arrows indicate that higher-order interpolation can eventually be used to obtain the centered part of the electric field at the z-edge. The thick horizontal and vertical arrows denote the normal components of the magnetic field at the zone faces. Owing to the divergence constraint, these field components are continuous across zone faces. They can, therefore, be obtained from the higher order facial reconstruction of the normal component of the magnetic field within each face. This provides higher order values of the x- and y-components of the magnetic field at the z-edge that minimize the dissipation terms.*

*Fig. 3 shows the same situation as in Fig. 2. However, it shows the situation after the four incoming states have interacted with each other. Four one-dimensional Riemann problems, shown by dashed lines, develop between the four pairs of incoming states. The resolved states from the one-dimensional HLL Riemann problems are shown by a superscript with a single star. The shaded*

*region depicts the strongly interacting state that arises when the four one-dimensional Riemann problems interact with one another. The strongly interacting state is shown by a superscript with a double star. We want to find the z-component of the electric field in the strongly interacting state. This gives us the z-component of the electric field at the z-edge, which is shown by the dot in this two-dimensional projection.*

*Figs. 4a to 4d show the four cases that are fully supersonic in both directions. The axes, in similarity variables, are shown in blue. The wave model is shown in black.*

*Figs. 4e to 4h show the four cases that are fully supersonic in only one of the two directions. The axes, in similarity variables, are shown in blue. The wave model is shown in black.*

*Fig. 5 shows the arrangement of the spatial nodes in the fourth order accurate RK-WENO algorithm for two space dimensions. The nodes within four abutting spatial zones are shown by the black dots. At fourth order, one-dimensional Gaussian quadrature requires the use of three quadrature points; we, therefore, see three nodes within each face. The red, double-sided arrows indicate the application of 1D Riemann solvers at the nodal points in the x-direction. The blue, double-sided arrows indicate the application of 1D Riemann solvers at the nodal points in the y-direction. The green dashed square at the right-upper vertex of zone (i,j) indicates the application of a 2D Riemann solver at the vertices of the mesh.*

*Fig. 6a and 6b are analogous to Fig. 2 because they show four zones in the xy-plane that come together at the z-edge of a three-dimensional mesh. Fig. 6a shows the four zones that surround a z-edge. It shows how the "θ" variables that are evaluated at the xz- and yz-faces can be used to form an effective "θ" at the z-edge of the mesh. This effective "θ" at the z-edge can then be used to lower the order of the edge-centered z-component of the electric field that is used in the update of the facial magnetic fields in the xz- and yz-faces. Like Fig. 2, Fig. 6b shows the inputs that go into the evaluation of the z-component of the electric field. The only difference from Fig. 2 is that we now have the option of making a high order evaluation (which uses all the high order WENO*

*reconstructions and interpolations as described in the text) which is superscripted with "HO"; and a low order (first order) evaluation which is superscripted with "LO".*

*Fig. 7 is analogous to Fig. 1 because it shows the components of the magnetic field in the faces of the mesh. The difference from Fig. 1 is that within each face we now have a high order component which is superscripted with "HO"; and a low order component which is superscripted with "LO". Both components in each face have been advanced in time using a forward Euler scheme with a timestep Δt. Both the components within each face will be used for the PCP update.*

*Fig. 8) We present order of accuracy plots for the MHD and RMHD Vortex problems. We show $L_1$ and $L_\infty$ errors for the x-momentum and x-magnetic field on a log scale versus the zone size. The top panel shows the accuracy plots for the MHD Vortex, and the bottom panel shows the accuracy plots for the RMHD Vortex problem. Both the MHD and RMHD vortex problems were run on a periodic domain that spans [-10,10]×[-10,10]. To facilitate comparison with Balsara (2004), the MHD vortex was run on $100^2$, $200^2$ and $400^2$ zone meshes. The RMHD vortex was run on $64^2$, $128^2$, $256^2$ and $512^2$ zone meshes to facilitate comparison with Balsara and Kim (2017). The top-left corner shows the slopes for different orders.*

*Fig. 9) We show plots of the magnetic energy distribution for the MHD-Vortex problem (obtained after one advection time). Following Leidi et al. (2022) we used $64^2$ zone mesh on a periodic domain that spans [-5,5]×[-5,5]. It is not useful to double the domain because we only focus on the maximum value of the magnetic energy. We normalize the obtained magnetic energy by the maximum initial magnetic energy. The top panel shows the distribution obtained from the multidimensional LLF Riemann solver version of the scheme, and the bottom panel shows the results obtained from the multidimensional HLL Riemann solver version of the scheme. Second to sixth order simulations have been presented.*

*Fig. 10) MHD Blast Problem: Fig. 10a shows the density profile, Fig. 10b shows the pressure profile, Fig. 10c shows the square of the magnitude of the velocity field vector, Fig. 10d shows the square of the magnitude of the magnetic field vector and Fig. 10e shows the value of the zone-centered θ that controls the amount of hybridization between high and low order schemes to obtain the PCP results. The profiles were obtained using a fourth-order scheme at time t=0.001 on a two-dimensional grid that consists of 400×400 zones. The plasma beta for this problem is β=2.51×10$^{-6}$. Finally, Fig. 10f shows the time evolution of the relative error in the total energy at three different mesh sizes: $400^2$, $800^2$, and $1600^2$.*

*Fig. 11) MHD Jet Problem: Fig. 11a shows the $\log_{10}$ of the density, Fig. 11b shows the $\log_{10}$ of the pressure, Fig. 11c shows the square of the magnitude of the magnetic field vector, and Fig. 11d shows the value of the zone-centered θ that controls the amount of hybridization between high and low order schemes to obtain the PCP results. The profiles were obtained at time t=0.002 using a sixth-order scheme on a two-dimensional grid that consists of 400×600 zones. The plasma beta for this problem is β=10$^{-4}$.*

*Fig. 12) RMHD Blast Problem: Fig. 12a shows the density profile, Fig. 12b shows the pressure profile, Fig. 12c shows the magnitude-squared of the Lorentz-velocity vector, Fig. 12d shows the magnitude-squared of the magnetic field vector and Fig. 12e shows the value of the zone-centered θ that controls the amount of hybridization between high and low order schemes to obtain the PCP results. The profiles were obtained using a fifth-order scheme at time t=4 on a two-dimensional grid that consists of 400×400 zones. The plasma beta for this problem is β=2.5×10$^{-6}$. Finally, Fig. 12f shows the time evolution of the relative error in the total energy at three different mesh sizes: $400^2$, $800^2$, and $1600^2$.*

*Fig. 13) RMHD Jet Problem: Fig. 13a shows the $\log_{10}$ of the density, Fig. 13b shows the $\log_{10}$ of the pressure, Fig. 13c shows the square of the magnitude of the magnetic field vector and Fig. 13d shows the value of the zone-centered θ that controls the amount of hybridization between*

*high and low order schemes to obtain the PCP results. The profiles were obtained at time t=30 using a fourth-order scheme on a two-dimensional grid that consists of 720×750 zones. The plasma beta for this problem is β=$10^{-3}$.*

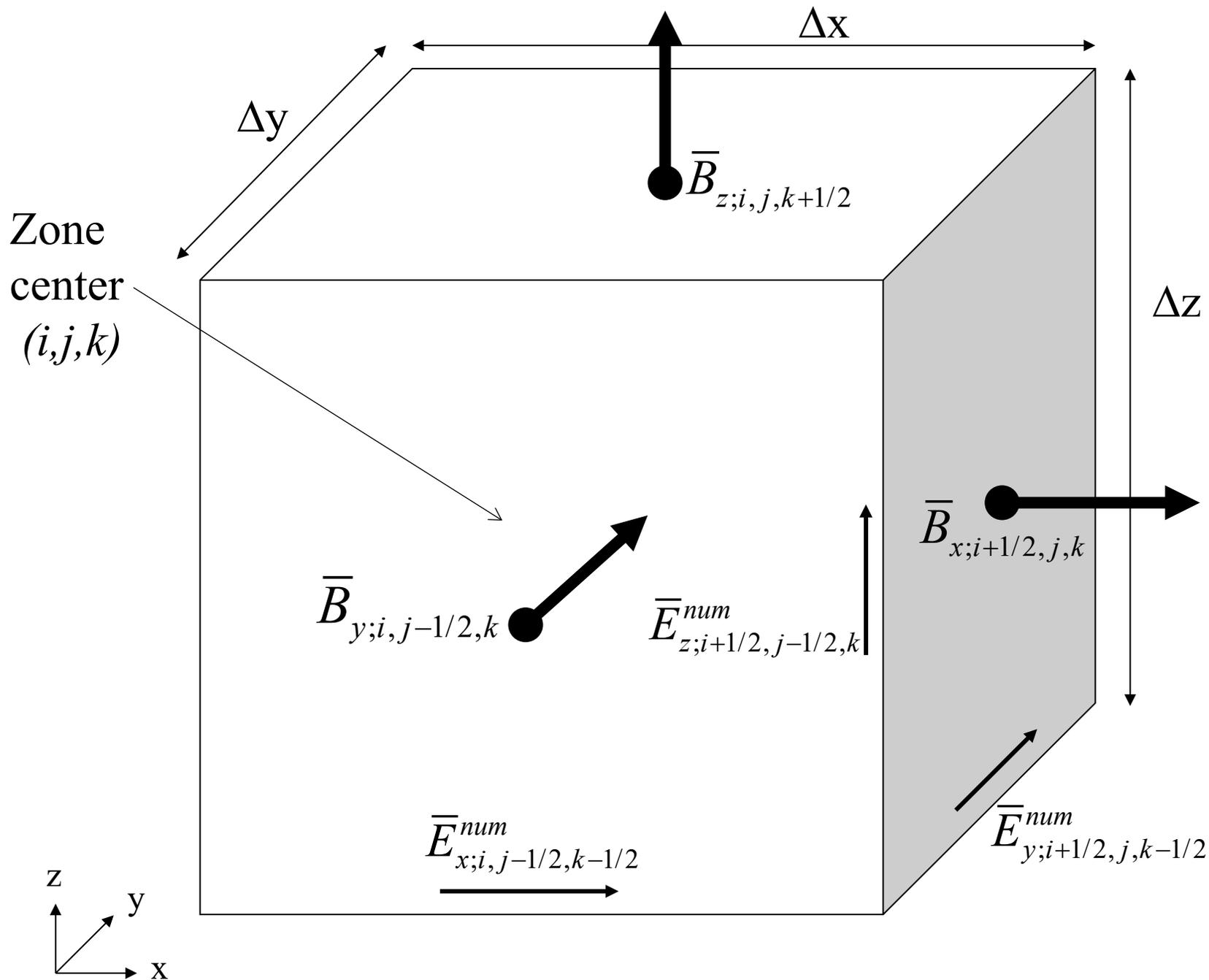

*Fig. 1 Most astrophysics and space physics use the Yee discretization shown here. The figure shows that the primal variables of the magnetic field, given by the normal components of the magnetic field, are facially-collocated. The components of the primal magnetic field vector are shown by the thick arrows. The overbars on the magnetic field components indicate that these are facially averaged. They undergo an update from the induction equation. The edge-collocated electric fields, which are used for updating the facial magnetic induction components, are shown by the thin arrows close to the appropriate edge. They too have overbars to indicate that they are edge-averaged. The superscript "num" for the electric field components indicates that they are multidimensionally stabilized and, therefore, suitable for use in a numerical scheme.*

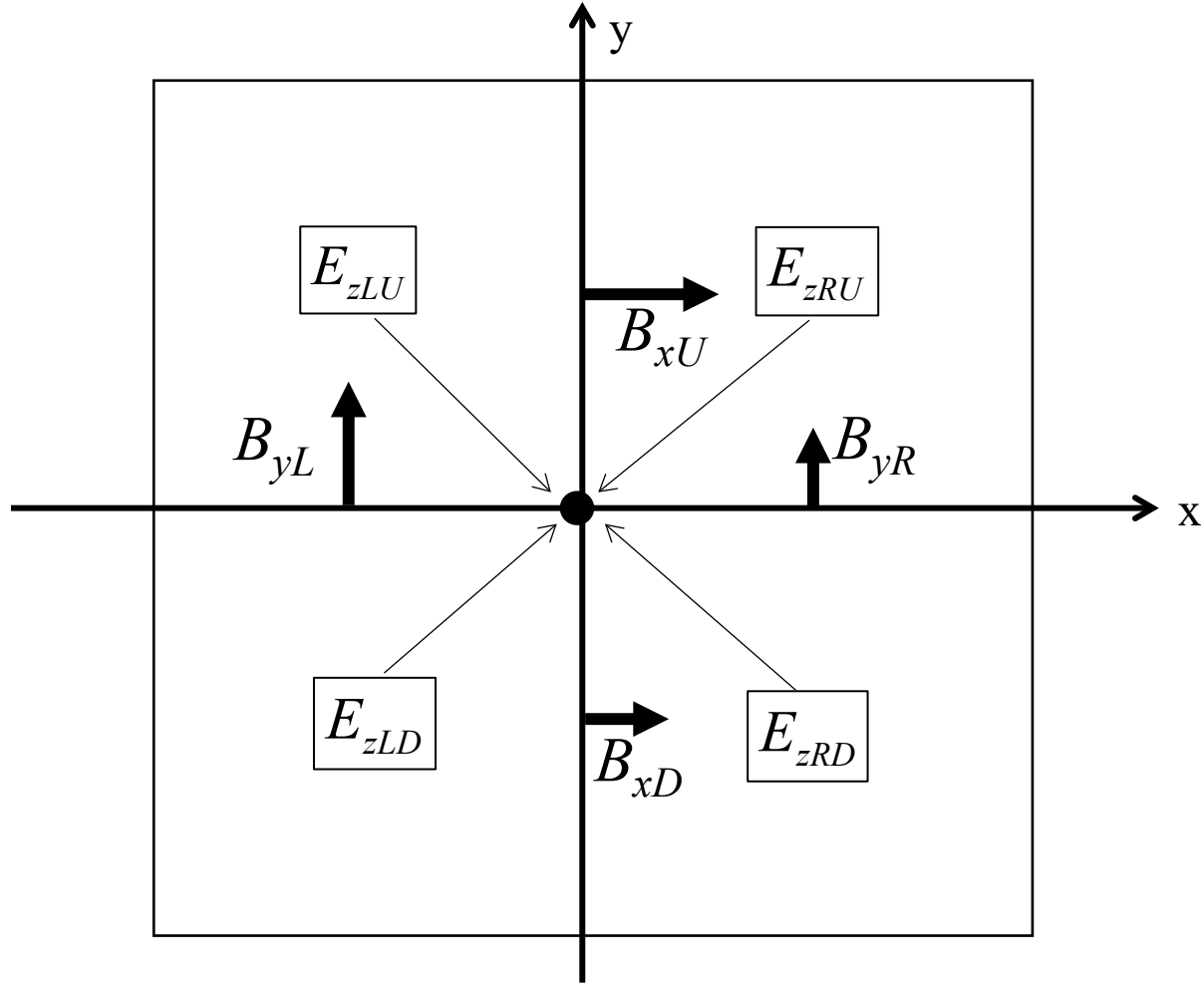

Fig. 2 shows four zones in the xy-plane that come together at the z-edge of a three-dimensional mesh. Since the mesh is viewed from the top in plan view, the z-edge is shown by the black dot and the four abutting zones are shown as four squares. The four incoming states have subscripts given by "RU" for right-upper; "LU" for left-upper; "LD" for left-down and "RD" for right-down. Fig 2 shows the situation before the states start interacting via four one-dimensional and one multidimensional Riemann problems. The thin oblique arrows indicate that higher-order interpolation can eventually be used to obtain the centered part of the electric field at the z-edge. The thick horizontal and vertical arrows denote the normal components of the magnetic field at the zone faces. Owing to the divergence constraint, these field components are continuous across zone faces. They can, therefore, be obtained from the higher order facial reconstruction of the normal component of the magnetic field within each face. This provides higher order values of the x- and y-components of the magnetic field at the z-edge that minimize the dissipation terms.

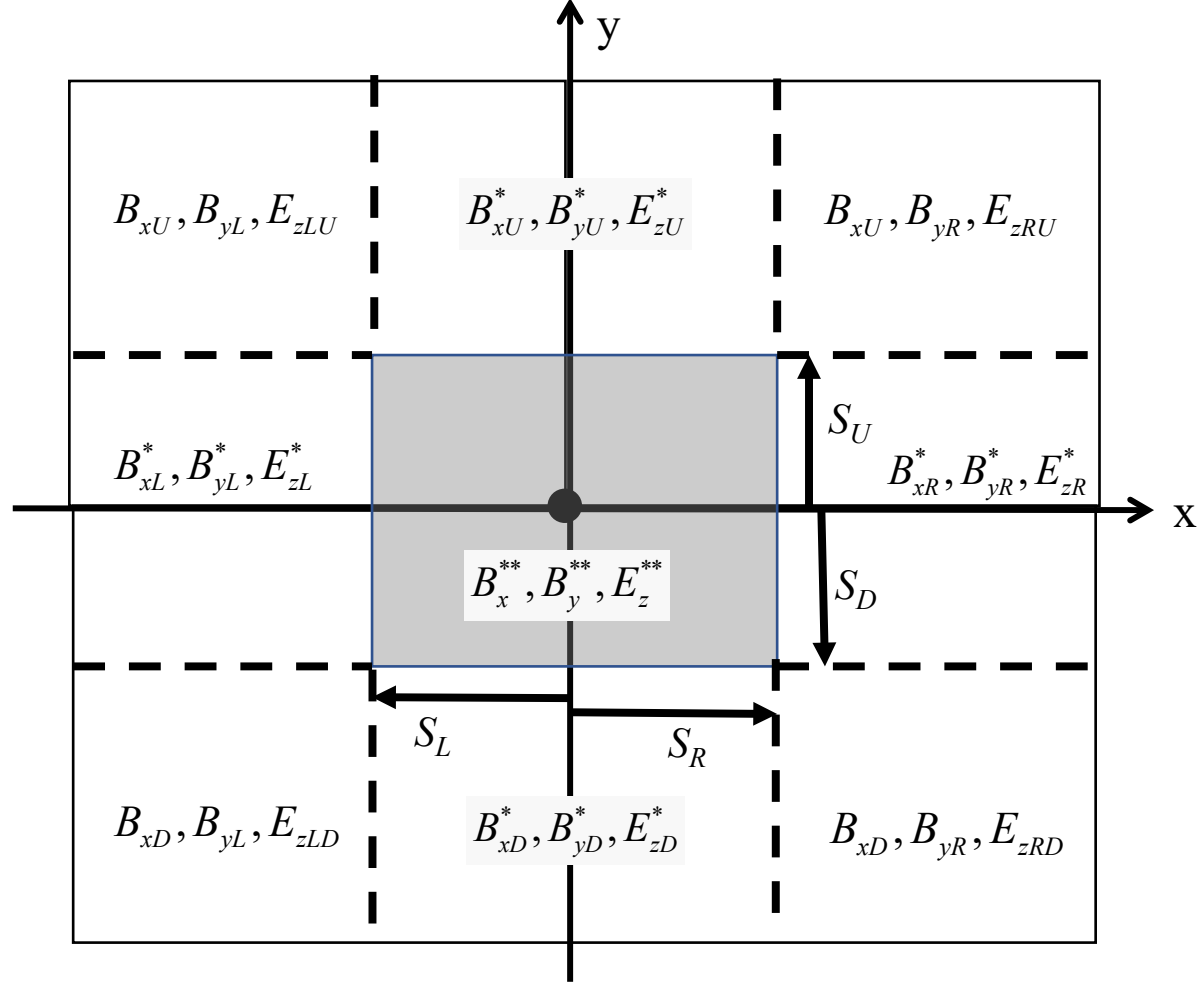

*Fig. 3 shows the same situation as in Fig. 2. However, it shows the situation after the four incoming states have interacted with each other. Four one-dimensional Riemann problems, shown by dashed lines, develop between the four pairs of incoming states. The resolved states from the one-dimensional HLL Riemann problems are shown by a superscript with a single star. The shaded region depicts the strongly interacting state that arises when the four one-dimensional Riemann problems interact with one another. The strongly interacting state is shown by a superscript with a double star. We want to find the z-component of the electric field in the strongly interacting state. This gives us the z-component of the electric field at the z-edge, which is shown by the dot in this two-dimensional projection.*

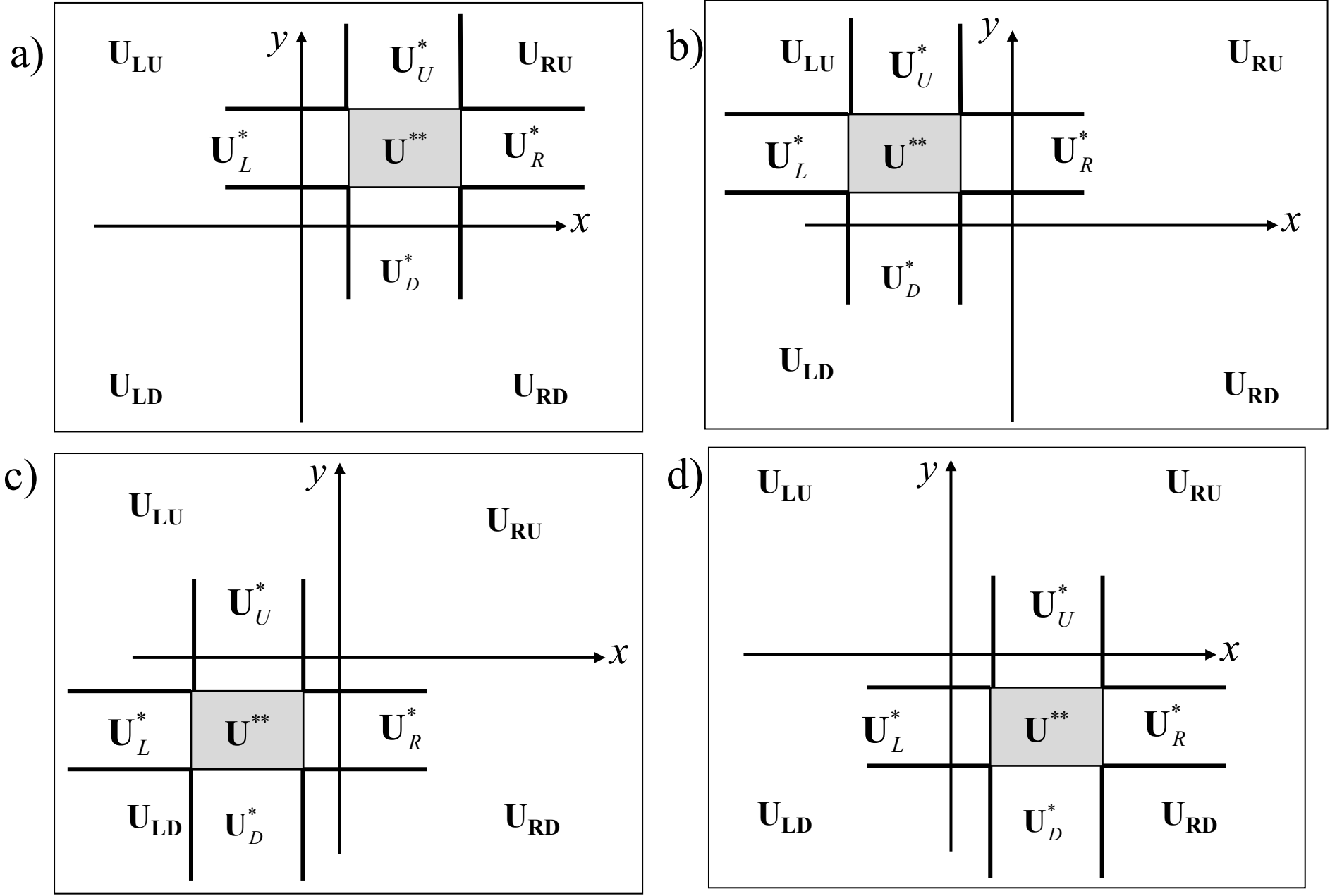

*Figs. 4a to 4d show the four cases that are fully supersonic in both directions. The axes, in similarity variables, are shown in blue. The wave model is shown in black.*

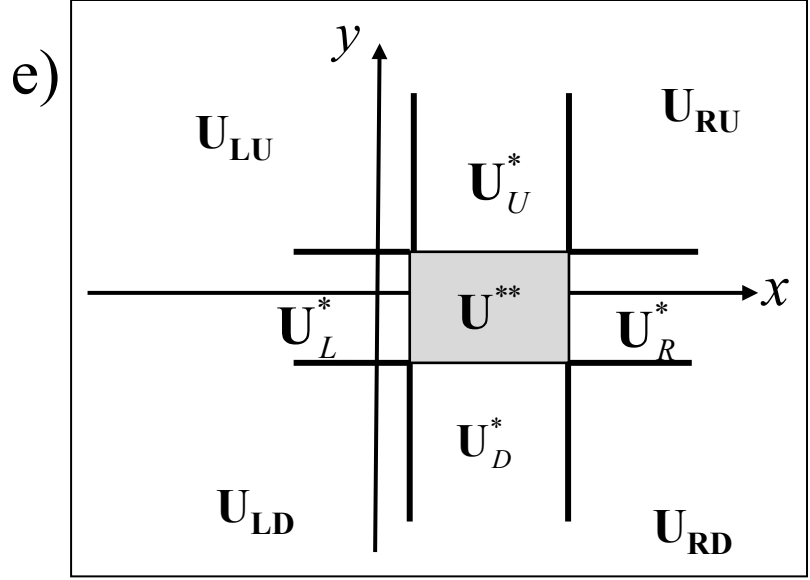
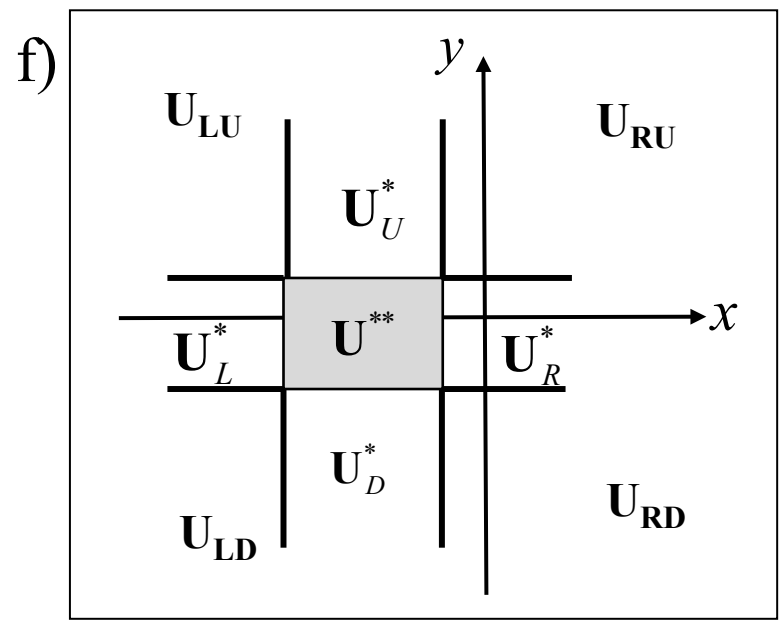
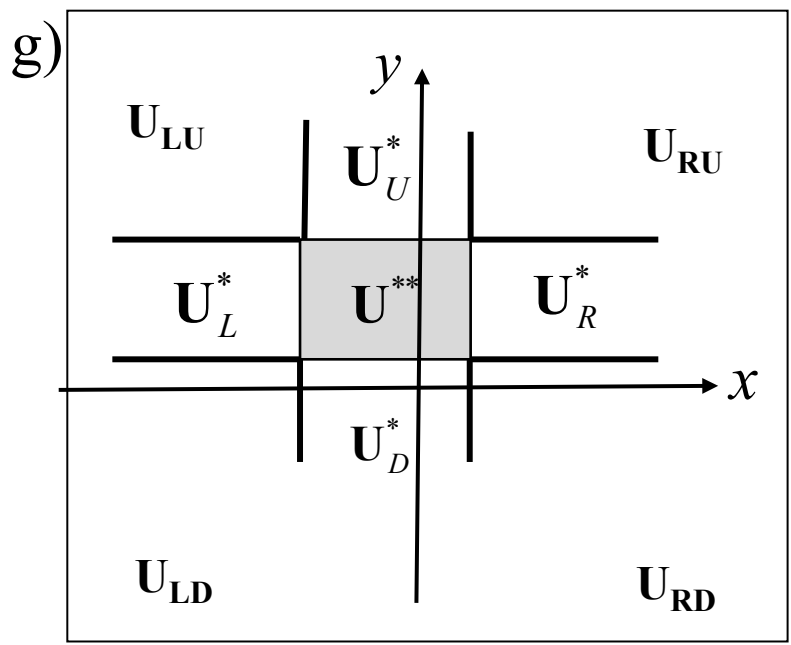
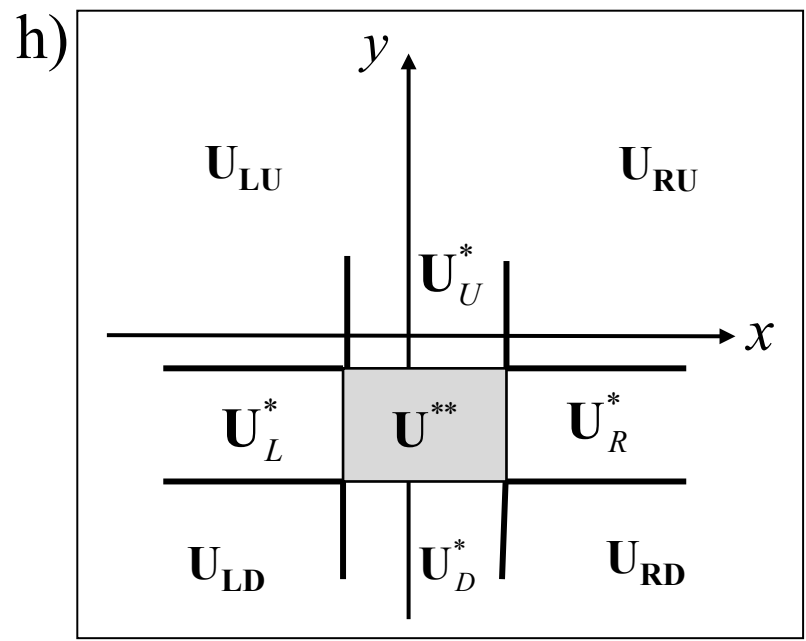

*Figs. 4e to 4h show the four cases that are fully supersonic in only one of the two directions. The axes, in similarity variables, are shown in blue. The wave model is shown in black.*

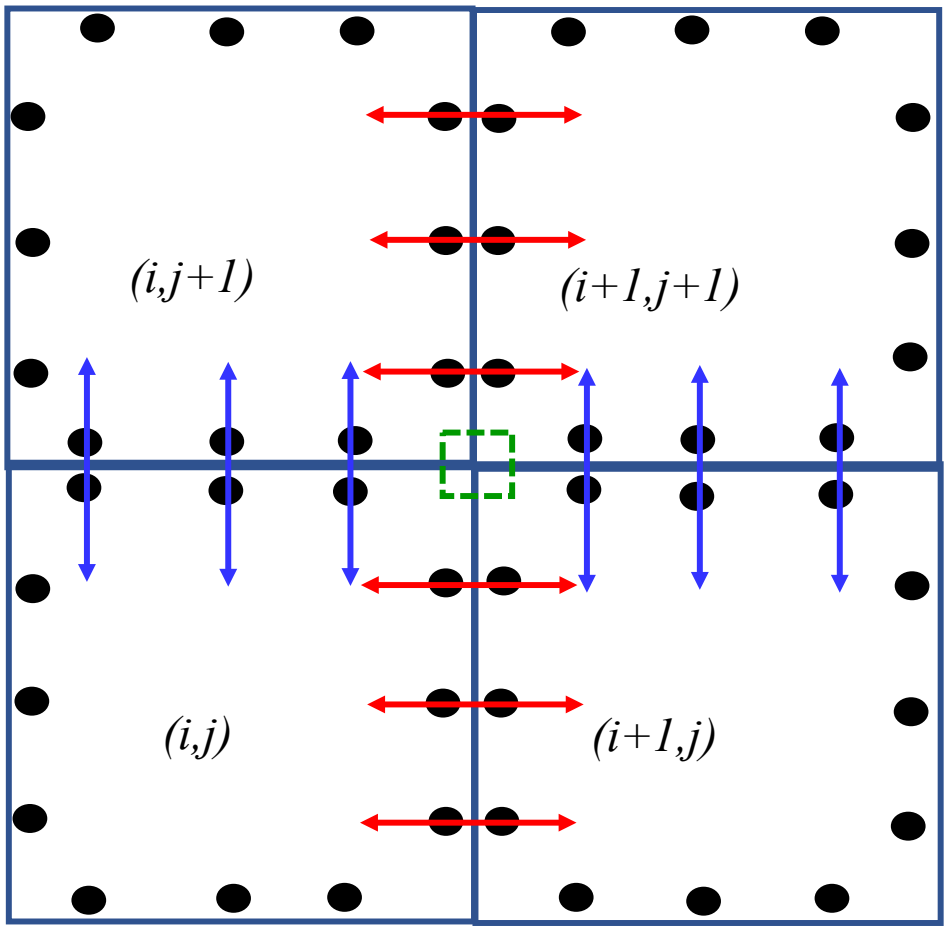

Fig. 5 shows the arrangement of the spatial nodes in the fourth order accurate RK-WENO algorithm for two space dimensions. The nodes within four abutting spatial zones are shown by the black dots. At fourth order, one-dimensional Gaussian quadrature requires the use of three quadrature points; we, therefore, see three nodes within each face. The red, double-sided arrows indicate the application of 1D Riemann solvers at the nodal points in the x-direction. The blue, double-sided arrows indicate the application of 1D Riemann solvers at the nodal points in the y-direction. The green dashed square at the right-upper vertex of zone (i,j) indicates the application of a 2D Riemann solver at the vertices of the mesh.

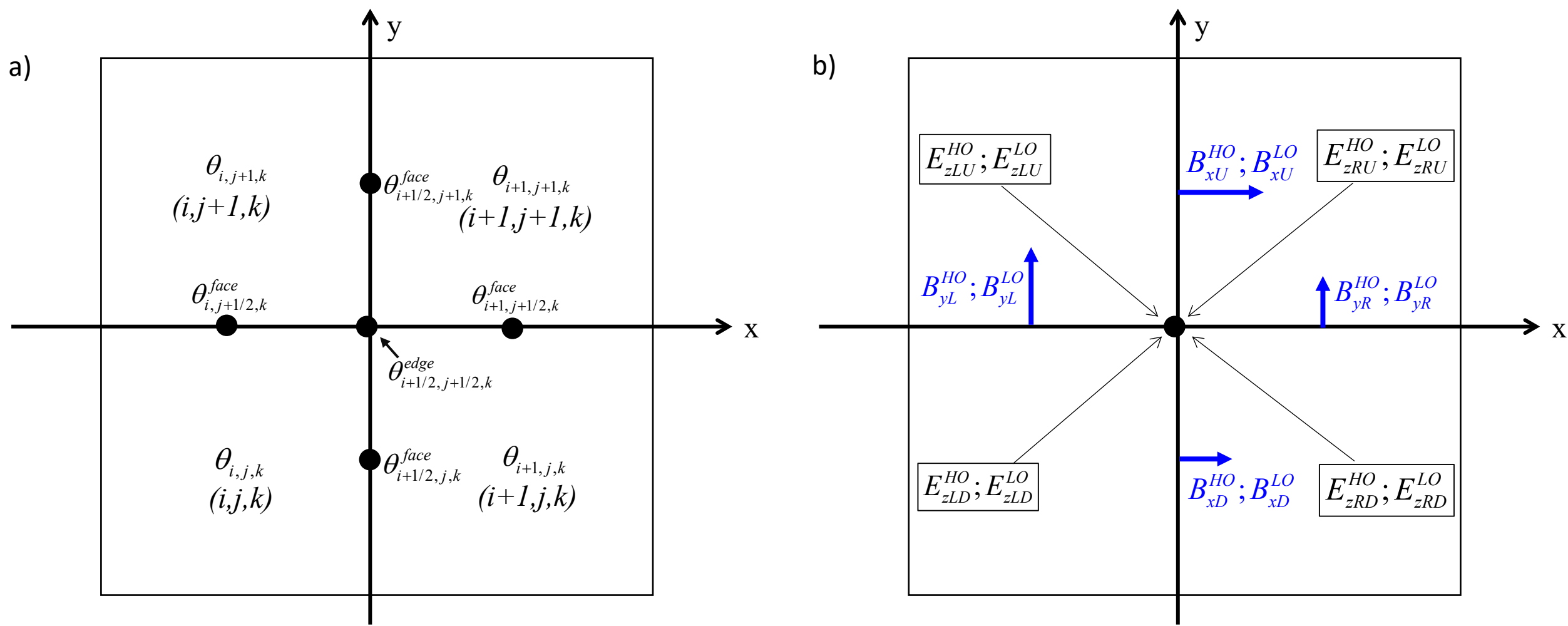

Fig. 6a and 6b are analogous to Fig. 2 because they show four zones in the xy-plane that come together at the z-edge of a three-dimensional mesh. Fig. 6a shows the four zones that surround a z-edge. It shows how the "θ" variables that are evaluated at the xz- and yz-faces can be used to form an effective "θ" at the z-edge of the mesh. This effective "θ" at the z-edge can then be used to lower the order of the edge-centered z-component of the electric field that is used in the update of the facial magnetic fields in the xz- and yz-faces. Like Fig. 2, Fig. 6b shows the inputs that go into the evaluation of the z-component of the electric field. The only difference from Fig. 2 is that we now have the option of making a high order evaluation (which uses all the high order WENO reconstructions and interpolations as described in the text) which is superscripted with "HO"; and a low order (first order) evaluation which is superscripted with "LO".

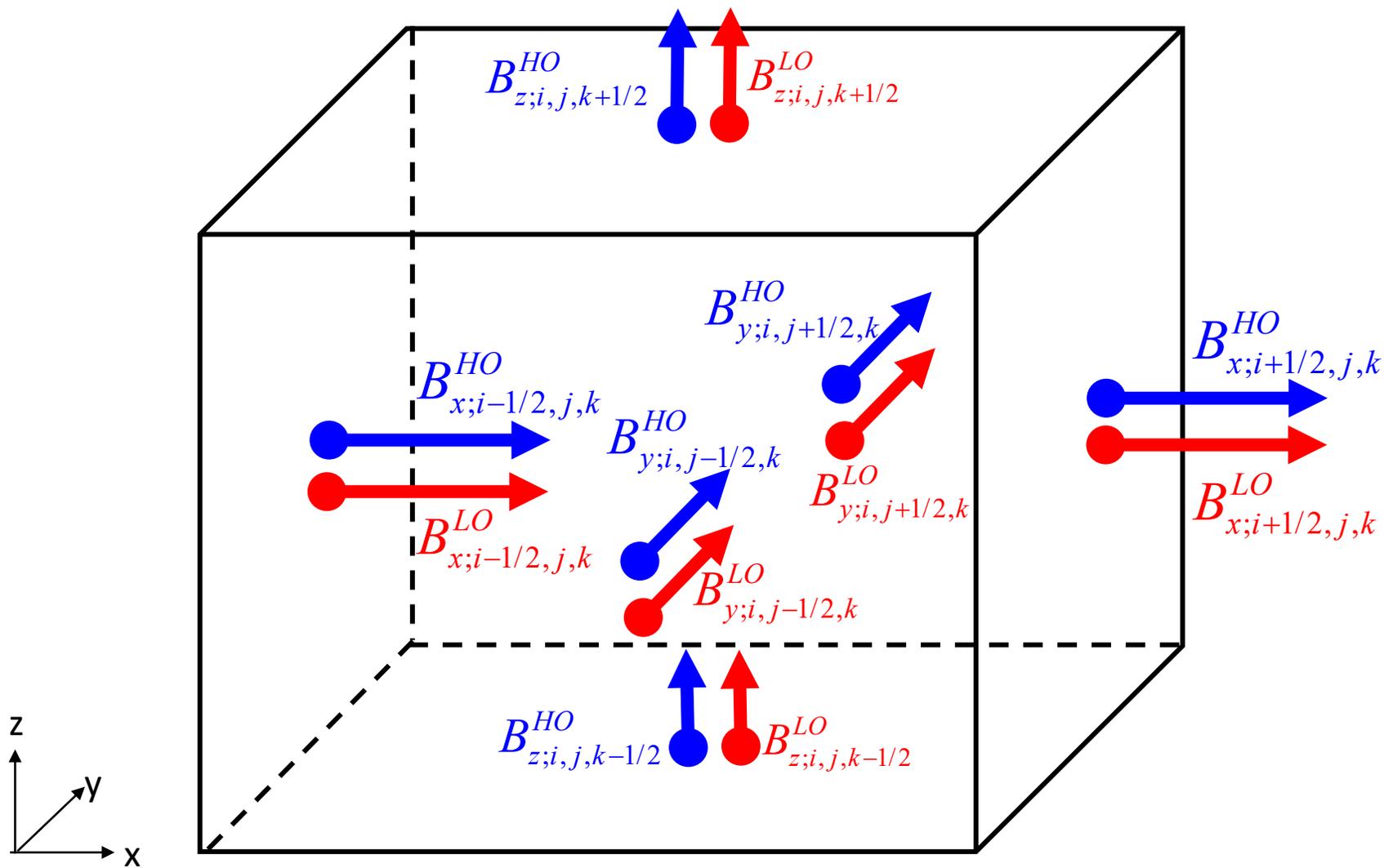

Fig. 7 is analogous to Fig. 1 because it shows the components of the magnetic field in the faces of the mesh. The difference from Fig. 1 is that within each face we now have a high order component which is superscripted with "HO"; and a low order component which is superscripted with "LO". Both components in each face have been advanced in time using a forward Euler scheme with a timestep $\Delta t$. Both the components within each face will be used for the PCP update.

MHD:

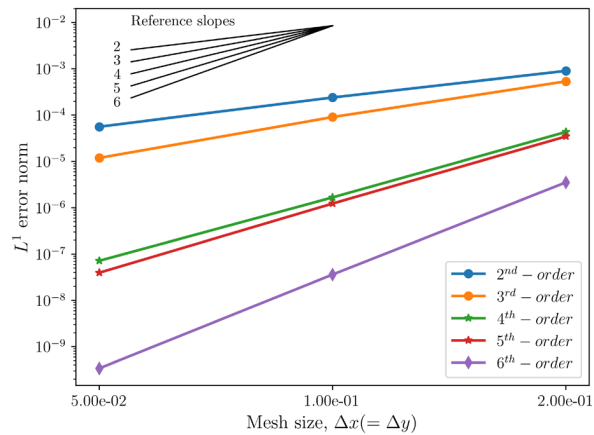 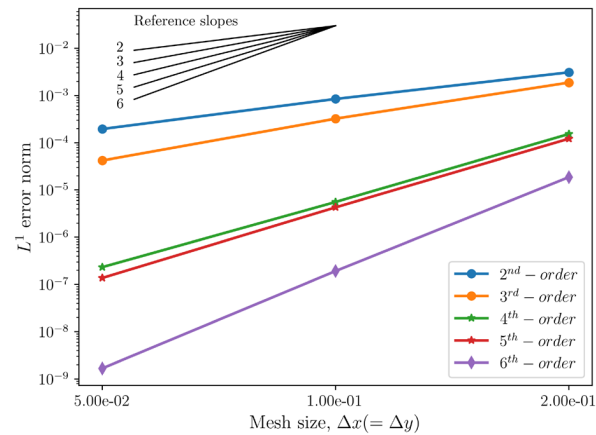 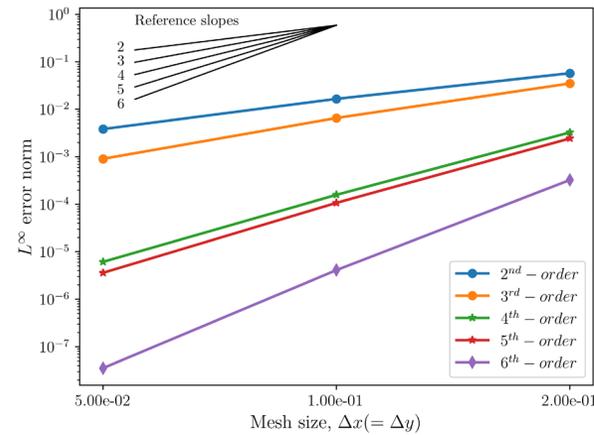 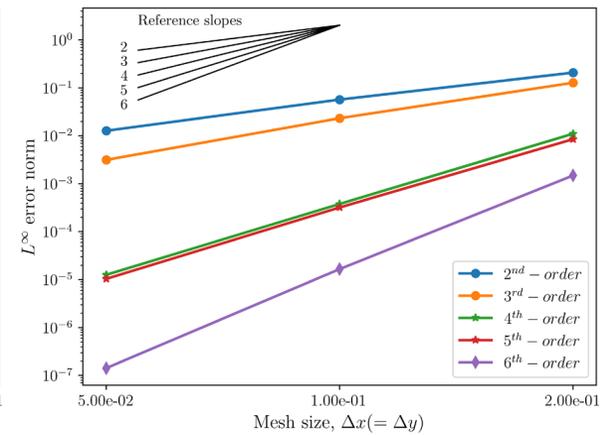

a) $L_1 - error$, variable $= M_x$    b) $L_1 - error$, variable $= B_x$    c) $L_\infty - error$, variable $= M_x$    d) $L_\infty - error$, variable $= B_x$

RMHD:

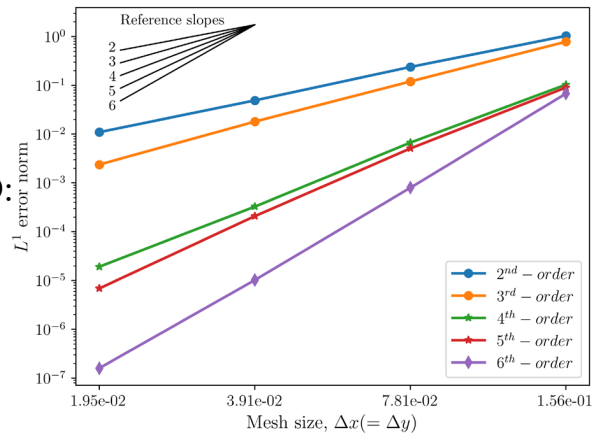 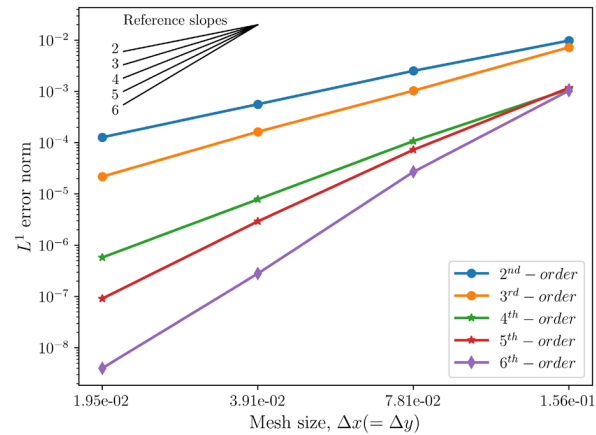 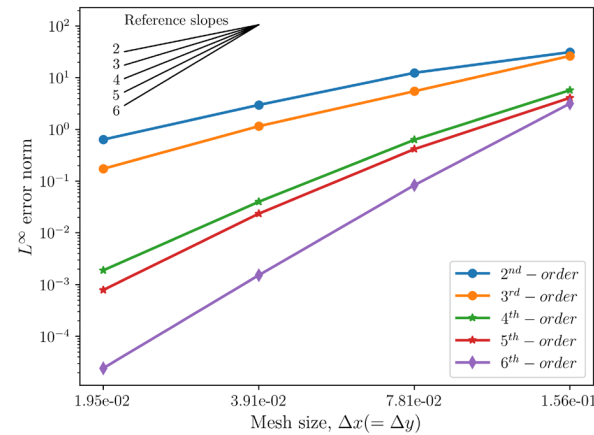 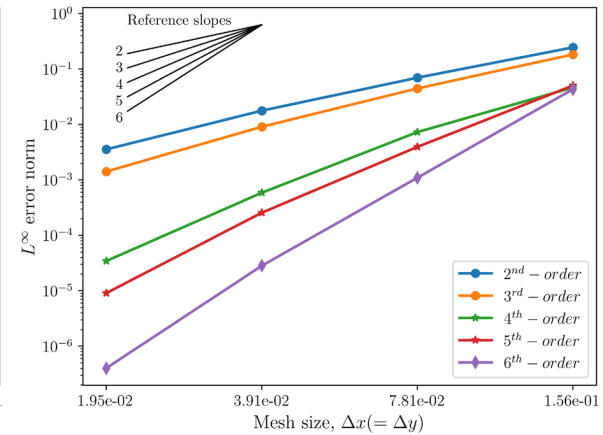

e) $L_1 - error$, variable $= M_x$    f) $L_1 - error$, variable $= B_x$    g) $L_\infty - error$, variable $= M_x$    h) $L_\infty - error$, variable $= B_x$

*Fig. 8) We present order of accuracy plots for the MHD and RMHD Vortex problems. We show $L_1$ and $L_\infty$ errors for the x-momentum and x-magnetic field on a log scale versus the zone size. The top panel shows the accuracy plots for the MHD Vortex, and the bottom panel shows the accuracy plots for the RMHD Vortex problem. Both the MHD and RMHD vortex problems were run on a periodic domain that spans [-10,10]×[-10,10]. To facilitate comparison with Balsara (2004), the MHD vortex was run on $100^2$, $200^2$ and $400^2$ zone meshes. The RMHD vortex was run on $64^2$, $128^2$, $256^2$ and $512^2$ zone meshes to facilitate comparison with Balsara and Kim (2017). The top-left corner shows the slopes for different orders.*

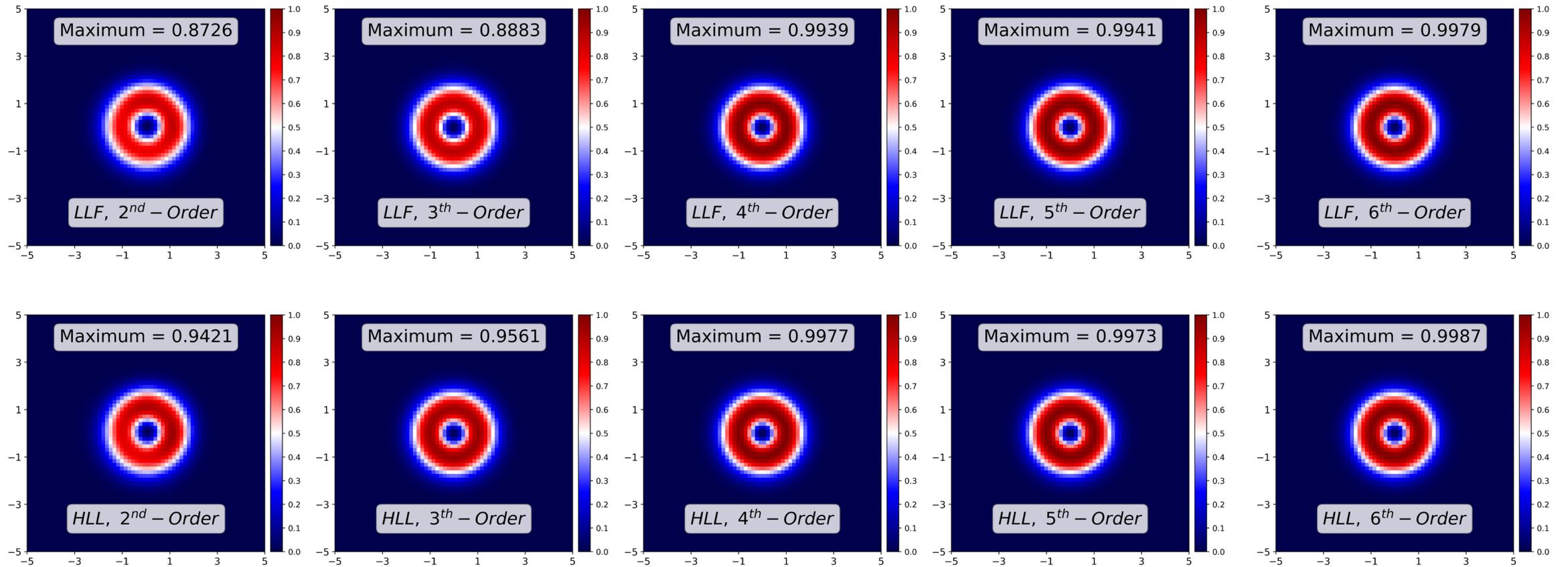

*Fig. 9) We show plots of the magnetic energy distribution for the MHD-Vortex problem (obtained after one advection time). Following Leidi et al. (2022) we used $64^2$ zone mesh on a periodic domain that spans $[-5,5] \times [-5,5]$. It is not useful to double the domain because we only focus on the maximum value of the magnetic energy. We normalize the obtained magnetic energy by the maximum initial magnetic energy. The top panel shows the distribution obtained from the multidimensional LLF Riemann solver version of the scheme, and the bottom panel shows the results obtained from the multidimensional HLL Riemann solver version of the scheme. Second to sixth order simulations have been presented.*

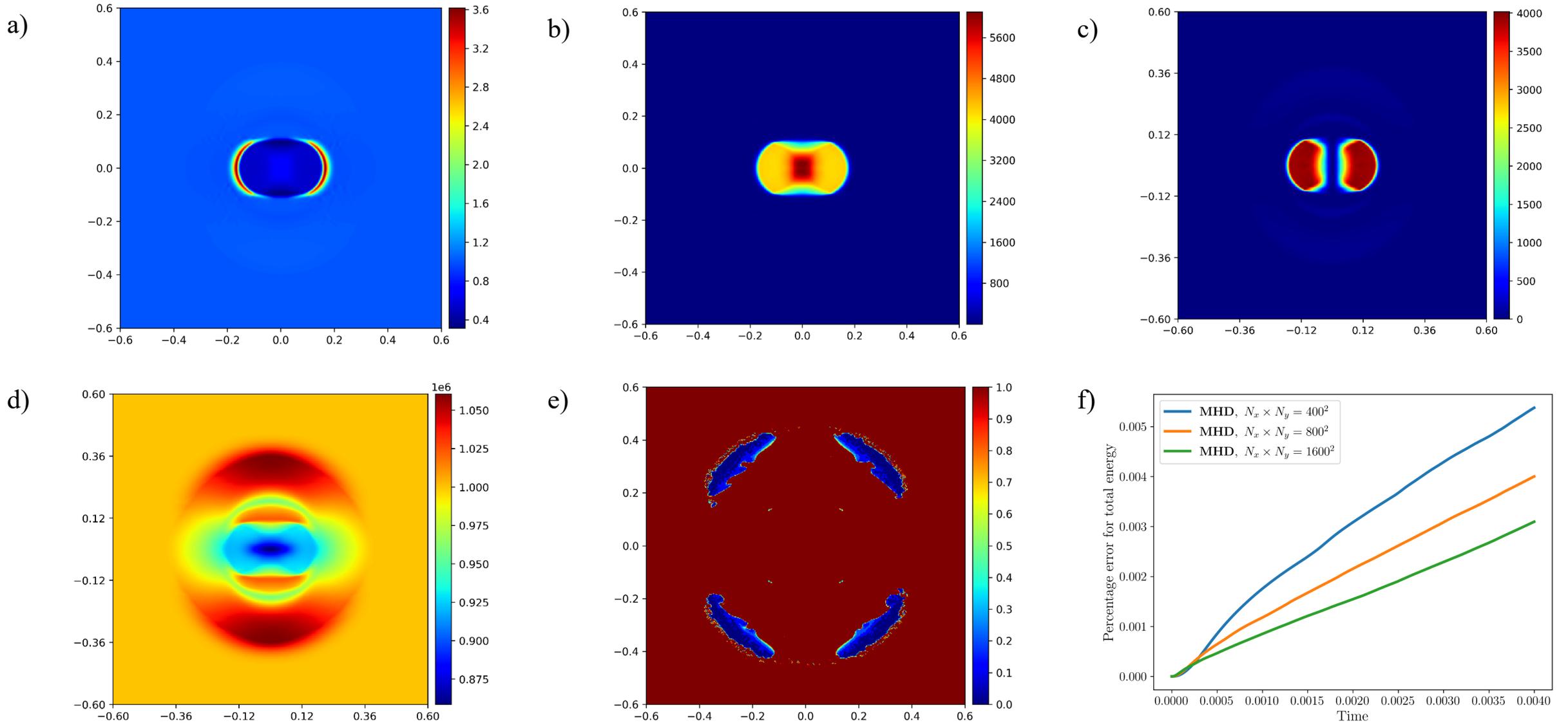

*Fig. 10) MHD Blast Problem: Fig. 10a shows the density profile, Fig. 10b shows the pressure profile, Fig. 10c shows the square of the magnitude of the velocity field vector, Fig. 10d shows the square of the magnitude of the magnetic field vector and Fig. 10e shows the value of the zone-centered θ that controls the amount of hybridization between high and low order schemes to obtain the PCP results. The profiles were obtained using a fourth-order scheme at time t=0.001 on a two-dimensional grid that consists of 400×400 zones. The plasma beta for this problem is β=2.51×10$^{-6}$. Finally, Fig. 10f shows the time evolution of the relative error in the total energy at three different mesh sizes: 400$^2$, 800$^2$, and 1600$^2$.*

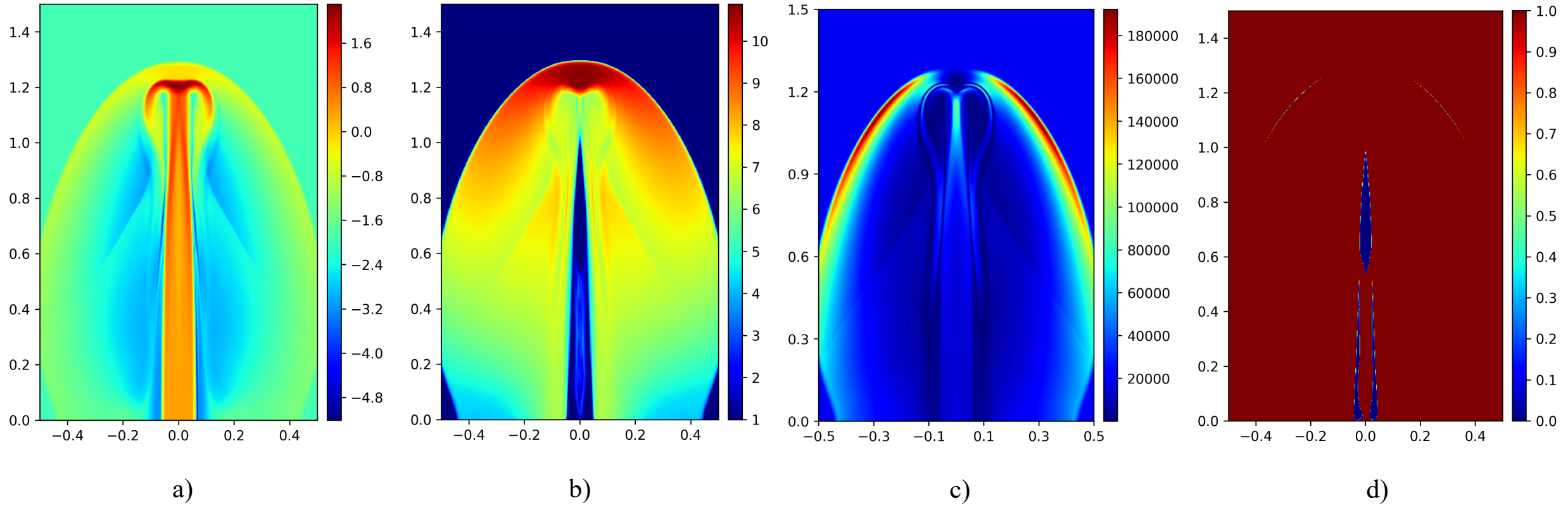

Fig. 11) MHD Jet Problem: Fig. 11a shows the $\log_{10}$ of the density, Fig. 11b shows the $\log_{10}$ of the pressure, Fig. 11c shows the square of the magnitude of the magnetic field vector, and Fig. 11d shows the value of the zone-centered $\theta$ that controls the amount of hybridization between high and low order schemes to obtain the PCP results. The profiles were obtained at time $t=0.002$ using a sixth-order scheme on a two-dimensional grid that consists of 400×600 zones. The plasma beta for this problem is $\beta=10^{-4}$.

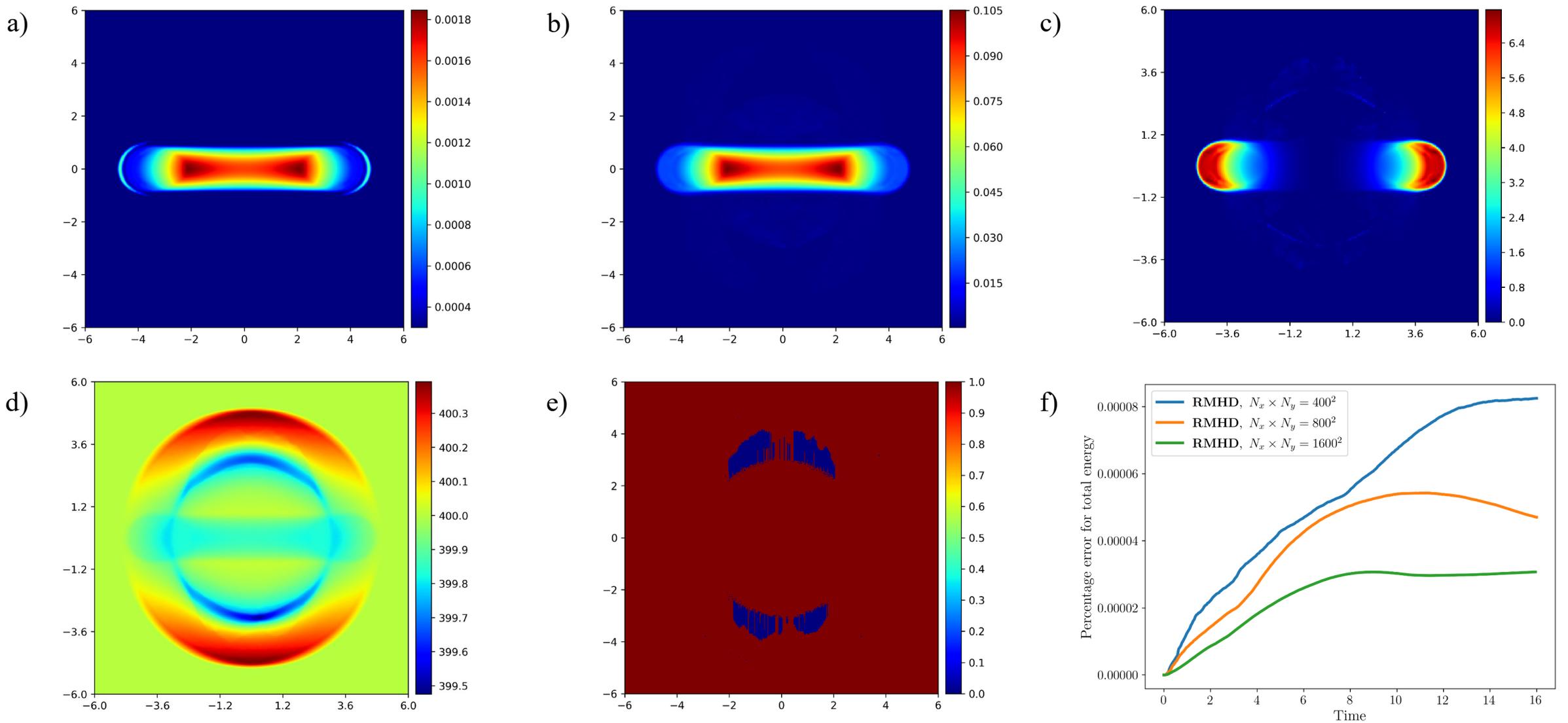

*Fig. 12) RMHD Blast Problem: Fig. 12a shows the density profile, Fig. 12b shows the pressure profile, Fig. 12c shows the magnitude-squared of the Lorentz-velocity vector, Fig. 12d shows the magnitude-squared of the magnetic field vector and Fig. 12e shows the value of the zone-centered θ that controls the amount of hybridization between high and low order schemes to obtain the PCP results. The profiles were obtained using a fifth-order scheme at time t=4 on a two-dimensional grid that consists of 400×400 zones. The plasma beta for this problem is β=2.5×10⁻⁶. Finally, Fig. 12f shows the time evolution of the relative error in the total energy at three different mesh sizes: 400², 800², and 1600².*

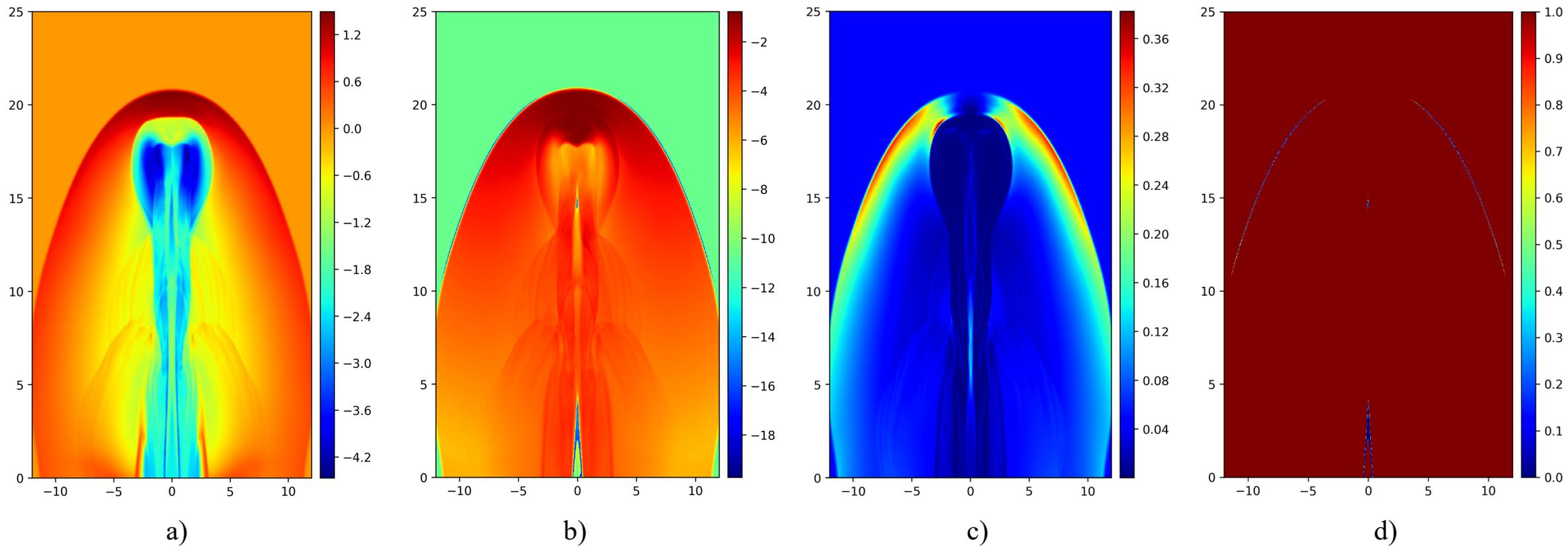

*Fig. 13) RMHD Jet Problem: Fig. 13a shows the $\log_{10}$ of the density, Fig. 13b shows the $\log_{10}$ of the pressure, Fig. 13c shows the square of the magnitude of the magnetic field vector and Fig. 13d shows the value of the zone-centered θ that controls the amount of hybridization between high and low order schemes to obtain the PCP results. The profiles were obtained at time t=30 using a fourth-order scheme on a two-dimensional grid that consists of 720×750 zones. The plasma beta for this problem is $\beta=10^{-3}$.*